\documentclass{article}
\usepackage[latin9]{inputenc}
\usepackage[a4paper]{geometry}
\usepackage{color}
\usepackage{float}
\usepackage{enumitem}
\usepackage{amsmath}
\usepackage{amsthm}
\usepackage{amssymb}
\usepackage{stmaryrd}
\usepackage{graphicx}
\usepackage[authoryear]{natbib}

\usepackage{dsfont} 
\usepackage{subcaption}
\newcommand{\lyxdot}{.}
\usepackage{nicefrac}
\usepackage{epstopdf}

\setcitestyle{round}

\makeatletter

\theoremstyle{plain}
\newtheorem{proposition}{\protect\propositionname}
\theoremstyle{definition}
\newtheorem{example}{\protect\examplename}
\theoremstyle{plain}
\newtheorem{theorem}{\protect\theoremname}
\theoremstyle{definition}
\newtheorem*{example*}{\protect\examplename}
\theoremstyle{plain}
\newtheorem{lemma}{\protect\lemmaname}
\theoremstyle{remark}
\newtheorem{remark}{\protect\remarkname}

\providecommand{\examplename}{Example}
\providecommand{\lemmaname}{Lemma}
\providecommand{\propositionname}{Proposition}
\providecommand{\theoremname}{Theorem}
\providecommand{\remarkname}{Remark}

\usepackage[linesnumbered,ruled,vlined]{algorithm2e}
\makeatletter
\renewcommand{\@algocf@capt@plain}{above}
\makeatother

\usepackage[unicode=true,pdfusetitle,
 bookmarks=true,bookmarksnumbered=false,bookmarksopen=false,
 breaklinks=false,pdfborder={0 0 0},pdfborderstyle={},backref=false,colorlinks=true]
 {hyperref}
\hypersetup{
 linkcolor=blue,citecolor=blue}

\begin{document}

\title{Revisiting the balance heuristic for estimating normalising constants}

\author{F J Medina-Aguayo
\thanks{Department of Mathematics and Statistics, University of Reading Whiteknights, PO Box 220, Reading RG6 6AX, UK} \\
{f.j.medinaaguayo@gmail.com} 
\and R G Everitt \thanks{Department of Statistics, University of Warwick, Coventry CV4 7AL, UK}}

\maketitle

\begin{abstract}
Multiple importance sampling estimators are widely used for computing
intractable constants due to its reliability and robustness. The celebrated
balance heuristic estimator belongs to this class of methods and has
proved very successful in computer graphics. The basic ingredients
for computing the estimator are: a set of proposal distributions,
indexed by some discrete label, and a predetermined number of draws
from each of these proposals. However, if the number of available
proposals is much larger than the number of permitted importance points,
one needs to select, possibly at random, which of these distributions
will be used. The focus of this work lies within the previous context,
exploring some improvements and variations of the balance heuristic
via a novel extended-space representation of the estimator, leading
to straightforward annealing schemes for variance reduction purposes.
In addition, we also look at the intractable scenario where the proposal
density is only available as a joint function with the discrete label,
as may be encountered in problems where an ordering is imposed. For this case, we look at combinations of correlated unbiased
estimators which also fit into the extended-space representation and,
in turn, will provide other interesting solutions.
\end{abstract}

\section{Introduction}

Computing normalising constants of probability distributions is of
great importance in many fields, and doing so remains being a challenge in complex scenarios. For instance, Bayesian inference is performed
via posterior distributions on a set of parameters and given some
data; depending on the specifics, these distributions are usually
known up to a constant of proportionality. Estimating this constant
would provide an insight to the the marginal density for the observed
data, which is also known as evidence or marginal likelihood. Among
the available Monte Carlo methods, importance sampling is the
usual approach for tackling this task \citep{RobertNCasella_2013};
importance sampling not only produces an unbiased estimate of the normalising constant
of some target distribution, but also a weighted sample for approximating
expectations under such target.

Despite the flexibility in the method and its validity under fairly
mild assumptions, a naive implementation of importance sampling is rarely adequate
for addressing the problem at hand, resulting in high variance estimates. Variance reduction techniques are commonly
used for overcoming this problem, e.g. the use of control variates,
defensive importance sampling, or the use of a mixture of proposals (see, e.g.,\citealp{Hesterberg_1995,HeNOwen_2014}
and \citealp{OwenNZhou_2000} for an overview).
In this paper we revisit one of such techniques termed balance heuristic, firstly introduced in \citet{VeachNGuibas_1995}, that belongs
to the broader class of multiple importance sampling methods
\citep{OwenNZhou_2000,ElviraETal_2015,ElviraETal_2019}. As the name
suggests, multiple importance sampling introduces multiple proposal distributions that are
indexed by a discrete label and that are then used for computing the
importance weights; this contrasts with standard importance sampling where only a single
proposal is used and that typically is inadequate for describing the
desired target distribution. Multiple importance sampling has proved popular in the area of
computer graphics, specifically in rendering applications which is
the main motivation in the seminal work \citet{VeachNGuibas_1995}.
When implementing multiple importance sampling one needs to address two aspects: the set of
proposals to be used and the corresponding number of samples to be
obtained from each one of these distributions. These matters have been
explored before giving rise to new methodology such as population
Monte Carlo \citep{CappeETal_2004}, adaptive multiple importance sampling \citep{CornuetETal_2012},
and related modifications and improvements, e.g. \citet{DoucETal_2007,MartinoETal_2015,ElviraETal_2017,Sbert_2017,MartinoETal_2017,Delyon_2018,BorsosETal_2019}
to name a few. From a different perspective, the authors in \citet{RoyETEvangelou_2018}
have proposed selecting proposals via a space filling criterion.

In this work we discuss some improvements and variations to the balance heuristic
estimator, all of these in the context where the multiple proposals
are randomly selected using an underlying distribution on the discrete
labels attached to them. This means that first a proposal is selected
from a (possibly large) pool by sampling a label from some discrete
distribution, then a value from the selected proposal distribution
is drawn which lies on the same space of the unnormalized target for
which the normalising constant is not known. As discussed later, the
resulting estimator remains valid and contrasts with the Rao-Blackwellized
one (obtained using the marginal proposal for the variable of interest,
i.e. summing the joint proposal density over the discrete label),
in the sense that the former is very competitive when considering
the cost involved in the marginalisation of the latter. This comparison
was not totally explored in \citet{VeachNGuibas_1995}, where balance heuristic is
only investigated using a fixed set of proposals and number of samples.
Our work focuses on two specific situations: when the number of available
proposal distributions is much larger than the permitted number of
importance points, and when the proposal densities show a level of
intractability (as encountered in multi-target tracking \citep{JiangETal_2015,LiETal_2017,JiangETSingh_2018},
or when using multiple proposals in sequential Monte Carlo samplers
\citep{MoralETal_2006,LiETal_2016,EverittETal_2016}).

In Section \ref{sec:MIS_BH} we look at the first scenario, where
we compare resulting estimates from the balance heuristic and the Rao-Blackwellized
estimators. Also, we introduce
a novel extended-space representation for balance heuristic which in turn motivates
the implementation of annealed importance sampling, as in \citet{Neal_2001},
for variance reduction purposes. In Section \ref{sec:intractable_proposals},
we explore the case where the joint proposal for the variable of interest
and the discrete label is intractable (in a sense defined later).
In order to deal with this intractability we build on existing work
on combining unbiased estimators, as in \citet{GramacyETal_2010,NguyenETal_2015,OwenNZhou_2019}.
We then show that combining estimators and balance heuristic can be regarded as specific
instances of a general framework, which motivates new schemes in this
intractable scenario. We conclude with a discussion and some final remarks
in Section \ref{sec:Conclusion}.

\section{The balance heuristic and extensions\label{sec:MIS_BH}}

\subsection{Overview of balance heuristic}
Suppose we are interested in estimating the normalising constant of
a target distribution $\pi$ on some measurable space $\left(\mathcal{X},\mathcal{B}\left(\mathcal{X}\right)\right)$.
More precisely, assume $\pi$ has a density $\pi\left(x\right)$ with respect to 
a dominating measure $\mu$ (e.g. Lebesgue or counting measure) and
given by
\[
\pi\left(x\right)=\frac{\tilde{\pi}\left(x\right)}{Z},
\]
where $\tilde{\pi}\left(x\right)$ is the unnormalized density and
$Z=\int_{\mathcal{X}}\tilde{\pi}\left(x\right)\mu\left(dx\right)$.
Our aim is to obtain estimates of $Z$.

A straightforward approach for estimating $Z$ is to use importance
sampling via an importance or proposal distribution $q$ such
that $\pi$ is absolutely continuous with respect to $q$. The resulting importance sampling estimator is unbiased and given
by
\[
\widehat{Z}=\frac{1}{N}\sum_{n=1}^{N}\frac{\tilde{\pi}\left(X_{n}\right)}{q\left(X_{n}\right)},\qquad X_{n}\sim q\left(\cdot\right).
\]
However, despite the simplicity of importance sampling, a good estimate can only be
achieved if $q$ and $\pi$ are close in some sense \citep[see, e.g.,][]{RobertNCasella_2013}.
It is often the case that
more than one importance distribution would be needed, e.g. in the
presence of multimodality on the target; in such setting, one can
resort to multiple importance sampling as in \citet{VeachNGuibas_1995}
for obtaining better estimates. To do this, let $\left\{ q_{i}\right\} _{i=1}^{K}$
be a family of importance distributions with $N_{i}$ denoting the number of draws obtained from $q_{i}$, and consider $\left\{ \omega_{i}\right\} _{i=1}^{K}$
be a set of weight functions such that $0\leq\omega_{i}\left(x\right)\leq\sum_{k=1}^{K}\omega_{k}\left(x\right)=1$
for any $x\in\mathcal{X}$, the multiple importance sampling estimator for $Z$ is
\begin{equation}\label{eq:Z_MIS}
\widehat{Z}_{MIS}=\sum_{i=1}^{K}\frac{1}{N_{i}}\sum_{j=1}^{N_{i}}\omega_{i}\left(X_{ij}\right)\frac{\tilde{\pi}\left(X_{ij}\right)}{q_{i}\left(X_{ij}\right)},
\end{equation}
where $X_{ij}\sim q_{i}\left(\cdot\right)$ for each $j\in\left\llbracket 1,N_{i}\right\rrbracket$ and $i\in\left\llbracket 1,K\right\rrbracket$, using the notation $\left\llbracket 1,n\right\rrbracket =\left\{ 1,\dots,n\right\} $ for any $n\in\mathbb{N}$. The previous estimate is computed using a
total of $N=\sum_{i=1}^{K}N_{i}$ draws from $K$ different distributions,
specifically $\left\{ X_{ij}\right\} _{j=1}^{N_{i}}\stackrel{iid}{\sim}q_{i}\left(\cdot\right)$
for every $i\in\left\llbracket 1,K\right\rrbracket $. We also note
that the functions $\left\{ \omega_{i}\right\} _{i}$ not only serve
as weights for draws within each $q_{i}$, but also as weights across
the different available proposals. Much of the current multiple importance sampling literature
focuses on the case where the number of available proposals ($K$)
is at most the number of proposed importance points ($N$) and assumes that
the variables $N_{1:K}=\left(N_{1},.\dots,N_{K}\right)$ are fixed;
here we consider $K\gg N$ which allows the possibility of selecting
proposals at random from the pool of size $K$. As a consequence, the resulting
vector $N_{1:K}$ will be random rather than deterministic.

Before discussing in depth the balance heuristic estimator, which is a special case
of $\widehat{Z}_{MIS}$, we state the following result that follows
from straightforward calculations, its proof can be found in Appendix 1 and follows from either \citealp{VeachNGuibas_1995,OwenNZhou_2000}.
\begin{proposition}
\label{prop:Z_BH}For fixed $N_{1:K}$ and any choice of weight functions
$\left\{ \omega_{i}\right\} _{i=1}^{K}$ the estimator $\widehat{Z}_{MIS}$
is unbiased with respect to $Z$, i.e. 
\[
E\left(\widehat{Z}_{MIS}\mid N_{1:K}\right)=Z.
\]
\end{proposition}
\begin{remark}
Using the Tower property, the previous result also shows that $\widehat{Z}_{MIS}$
is unbiased when the variables $N_{1:K}$ are random. This will be
the setting considered throughout the rest of this work.
\end{remark}

The balance heuristic estimator from \citet{VeachNGuibas_1995} corresponds to a
specific instance of $\widehat{Z}_{MIS}$ using a specific set of
weight functions $\left\{ \omega_{i}\right\} _{i}$, these are defined
as follows for every $i\in\left\llbracket 1,K\right\rrbracket$
\begin{equation}
\omega_{i}^{BH}\left(x\right)=\frac{N_{i}q_{i}\left(x\right)}{\sum_{k=1}^{K}N_{k}q_{k}\left(x\right)}.\label{eq:weight_BH}
\end{equation}
The above weights lead to the following estimate of $Z$,
\begin{equation}
\widehat{Z}_{BH}=\sum_{i=1}^{K}\sum_{j=1}^{N_{i}}\frac{\tilde{\pi}\left(X_{ij}\right)}{\sum_{k=1}^{K}N_{k}q_{k}\left(X_{ij}\right)}.\label{eq:Z_BH}
\end{equation}
The choice in \eqref{eq:weight_BH} is closely related to the Rao-Blackwellized estimator, which is computed using a sample $\left\{ X_{n}\right\} _{n=1}^{N}$
obtained as follows. For each $n\in\left\llbracket 1,N\right\rrbracket $:
\begin{itemize}
\item Draw a label $L_{n}$ from an auxiliary distribution $\alpha$ on
the space $\left(\left\llbracket 1,K\right\rrbracket ,\mathcal{P}\left(\left\llbracket 1,K\right\rrbracket \right)\right)$;
\item Draw $X_{n}\mid L_{n}\sim q_{L_{n}}\left(\cdot\right)$ independently
from $L_{-n}$ and $X_{-n}$, using the notation 
\[
L_{-n}=(L_1,\dots,L_{n-1},L_{n+1},\dots,L_N).
\]
\end{itemize}
The Rao-Blackwellized estimator is the resulting importance sampling estimator considering the marginal
distribution of the sample $\left\{ X_{n}\right\} _{n}$ as importance
distribution, this is

\begin{equation}
\widehat{Z}_{RB}=\frac{1}{N}\sum_{n=1}^{N}\frac{\tilde{\pi}\left(X_{i}\right)}{\sum_{k=1}^{K}\alpha\left(k\right)q_{k}\left(X_{i}\right)},\qquad X_{n}\mid L_{n}\sim q_{L_{n}}\left(\cdot\right), L_{n}\sim\alpha\left(\cdot\right).\label{eq:Z_opt}
\end{equation}
The connection between balance heuristic and Rao-Blackwellized is clear by noting that $\widehat{Z}_{RB}$
can also be expressed as 
\begin{align*}
\widehat{Z}_{RB} & =\sum_{n=1}^{N}\omega_{L_{n}}^{RB}\left(X_{n}\right)\frac{\tilde{\pi}\left(X_{n}\right)}{N\alpha\left(L_{n}\right)q_{L_{n}}\left(X_{n}\right)},
\end{align*}
where
\begin{equation}
\omega_{l}^{RB}\left(x\right)=\frac{\alpha\left(l\right)q_{l}\left(x\right)}{\sum_{k=1}^{K}\alpha\left(k\right)q_{k}\left(x\right)},\qquad l\in\left\llbracket 1,K\right\rrbracket .\label{eq:weight_RB}
\end{equation}
Comparing \eqref{eq:weight_BH} and \eqref{eq:weight_RB} we observe they are equivalent if we set $\alpha\left(l\right)=N_{l}/N$
for each $l\in\left\llbracket 1,K\right\rrbracket $. Nevertheless,
the resulting estimators remain different in the way they are constructed.
For the balance heuristic, the variables $N_{1:K}$ can be drawn jointly from any
appropriate distribution; whereas in the Rao-Blackwellized case, the common auxiliary
distribution $\alpha$ for drawing the labels only leads to $N_{1:K}\sim Mult\left(N,\alpha_{1:K}\right)$,
where $\alpha_{l}=\alpha\left(l\right)$ for each $l\in\left\llbracket 1,K\right\rrbracket $.
Moreover, even if $N_{1:K}\sim Mult\left(N,\alpha_{1:K}\right)$ in
both settings, which is the case we consider throughout, the estimators will
have very different properties. We will revisit the Rao-Blackwellized estimator in
the next section, which in the literature is usually thought of an
\textit{optimal} estimator in a sense that is defined later.

We now provide qualitative bounds for the difference in relative variance of $\widehat{Z}_{BH}$ and $\widehat{Z}_{RB}$, provided the proposal densities can be bounded away from $0$ and $\infty$ (this easily holds e.g. when $\mathcal{X}$ is compact). The proof can be found in Appendix 1.

\begin{theorem}\label{thm:vars}
Suppose that there are positive constants $C_{-}$ and $C_{+}$ such that, for every $i\in \left\llbracket 1,K\right\rrbracket$, $\inf_{x\in\mathcal{X}} q_i(x) \geq C_{-}>0$ and $\sup_{x\in\mathcal{X}} q_i(x)\leq C_{+}<\infty$. Then, for any choice of $\alpha_{1:K}$, the  variance of  $\widehat{Z}_{BH}$ satisfies
\begin{align*}
-\frac{\tilde{C}_{-}}{N} \leq \mbox{var}\left( \widehat{Z}_{BH}/Z \right) - \mbox{var}\left( \widehat{Z}_{RB}/Z \right)  &\leq  \frac{\tilde{C}_{+}}{N^2}
\end{align*}
for some $\tilde{C}_{-},\tilde{C}_{+}>0$.
\end{theorem}

\begin{remark}
When $N=K=2$ and for $\alpha_1=\alpha_2=1/N$ we can take $\tilde{C}_{-}= 0$, as shown in \cite{ElviraETal_2019}; however, it is not clear whether this will hold in more general settings. Precise values for $\tilde{C}_{-}$ and $\tilde{C}_{+}$ are given in the proof.
\end{remark}

Recently, the authors from \citet{SbertETal_2016,SbertETal_2018}
discuss the choice of a deterministic $N_{1:K}$ for minimising the
variance of $\widehat{Z}_{MIS}$; the optimal choice for each $N_{i}$
depends on the following quantity
\[
\sigma_{i}^{2}=\mbox{var}\left(\omega_{i}\left(X\right)\frac{\tilde{\pi}\left(X\right)}{q_{i}\left(X\right)}\right),\qquad X\sim q_{i}\left(\cdot\right);
\]
which measures the variability of the contribution to the overall
estimate $\widehat{Z}_{MIS}$ from the proposal $q_{i}$. As a consequence,
the variance of $\widehat{Z}_{BH}$ when using a deterministic $N_{1:K}$
can also be minimized obtaining a provably better estimate than the
one resulting from $N_{l}\propto1$ for all $l\in\left\llbracket 1,K\right\rrbracket$.
Clearly, a downside to this optimal approach is the need to know or
at least estimate accurately the vector $\sigma_{1:K}^{2}$. It is
also worth pointing out the recent work in \citet{SbertNElvira_2019},
which discusses a generalisation of balance heuristic by introducing an extra variable
in \eqref{eq:Z_BH} for obtaining provably better estimates; however,
an optimal setting for this approach also relies on knowing $\sigma_{1:K}^{2}$.
Our work diverges from the aforementioned articles in the sense that
we do not attempt to optimise in terms of $N_{1:K}$ or its underlying
distribution, instead we only allow for $N_{1:K}$ to be random (distributed
according to $Mult\left(N,\alpha_{1:K}\right)$) and explore ideas
to reduce the variance of the resulting estimator.

Although the estimator in \eqref{eq:Z_BH} might not be optimal in
terms of variance (see \citealp[Theorem 1]{VeachNGuibas_1995} and
\citealp[Section 4]{SbertETal_2018}), it can still provide accurate
estimates with a moderate cost. In fact, when $K$ (the number of
importance distributions) is much larger than $N=\sum_{i=1}^{K}N_{i}$
(the total number of importance points), the balance heuristic estimator is less
computationally expensive since many of the variables $N_{1:K}$ will
be equal to zero. The cost of implementing such scheme is at most
$\mathcal{O}\left(N^{2}\right)$ compared to the optimal method with
cost $\mathcal{O}\left(NK\right)$. This becomes clear by re-expressing
\eqref{eq:Z_BH} as
\begin{align}
\widehat{Z}_{BH} & =\sum_{n=1}^{N}\frac{\tilde{\pi}\left(X_{n}\right)}{\sum_{m=1}^{N}q_{L_{m}}\left(X_{n}\right)},\qquad X_{n}\mid L_{n}\sim q_{L_{n}}\text{\ensuremath{\left(\cdot\right)}}, L_{n}\sim\alpha\left(\cdot\right).\label{eq:Z_BH_alt}
\end{align}
Define $K_{eff}=\sum_{l=1}^{K}\mathds{1}\left(L_{i}=l\text{ for some }i\in\left[1:N\right]\right)$
as the number of effective labels sampled from the set $\left\llbracket 1,K\right\rrbracket $,
the actual computational cost of the balance heuristic estimator is $\mathcal{O}\left(NK_{eff}\right)$,
noting that $K_{eff} \leq \min\left\{ K,N\right\}$. Therefore, provided that the variance for the balance heuristic estimator is not much larger than that of the Rao-Blackwellized estimator, we would expect a better performance from the former for the same computational cost. We now present an illustrative running example.
\begin{example}[Running example]
\label{exa:Example1}Let $\pi$ be Gaussian target $\pi\left(x\right)\propto\exp\left\{ -0.5 x^2\right\} $, and for simplicity we use $Z=1$. Also consider
a set of proposal distributions $\left\{ q_{l}\right\} _{l=1}^{K}$
with corresponding probabilities of being selected $\left\{ \alpha_{l}\right\} _{l=1}^{K}$, namely the associated densities are
\[
q_{l}\left(x\right)\propto\exp\left\{ -\frac{\left(x-\mu_{l}\right)^{2}}{4}\right\} \quad\text{and}\quad\alpha_{l}=BetaBinom\left(l\mid K,m,s\right),
\]
where $\mu_{l}=\left(\mu_{max}-\mu_{min}\right)\left(l-1\right)/ \left(K-1\right)+\mu_{min}$
and $BetaBinom\left(\cdot\mid K,m,s\right)$ denotes the beta-binomial
distribution of parameters $m>0$, $s>0$ with density
\[
BetaBinom\left(l\mid K,m,s\right)=\binom{K}{l}\frac{B\left(l+sm,K-l+s\left(1-m\right)\right)}{B\left(sm,s\left(1-m\right)\right)},\qquad l\in\left\llbracket 1,K\right\rrbracket.
\]
The specification above implies that the distributions are equally
spaced between $\mu_{min}$ and $\mu_{max}$, and the weights associated
to each proposal within the set are given by the beta-binomial distribution
of parameters $m$ and $s$. In this respect, the parameter $m>0$
controls the symmetry of the distribution around its mean $Km$, whereas
$s>0$ is a concentration parameter. Two particular cases are $BetaBinom\left(K,m,s=\infty\right)$ which results in a $Binom\left(K-1,m\right)$
and $BetaBinom\left(K,m=0.5,s=2\right)$ resulting in $Unif\left\llbracket 1,K\right\rrbracket $.

Figure \ref{fig:Fig1} compares the estimators $\widehat{Z}_{BH}$
and $\widehat{Z}_{RB}$ for the standard Gaussian target and
for different sets of proposals. When
$K$ is small in comparison to $N$ the probability that $K_{eff}$ equals $K$ is very
close to 1; however, as $K$ increases this probability decreases
and becomes zero for $K>N$. Hence, computing the estimator
$\widehat{Z}_{BH}$ is much cheaper than computing $\widehat{Z}_{RB}$
when $K_{eff}<K$, which results in smaller variance for equivalent
computational costs. This is confirmed by plots in Subfigre (c) where the number of
importance points used for computing $\widehat{Z}_{RB}$ has to be
reduced dramatically. Observe that balance heuristic provides sensible
estimates even when $K=30,000$ or when the proposal is more concentrated, and is competitive
to $\widehat{Z}_{RB}$ when $K$ is small.

\begin{figure}[!ht]{
\begin{subfigure}[l]{0.48\linewidth}
\centering
\includegraphics[scale=0.7]{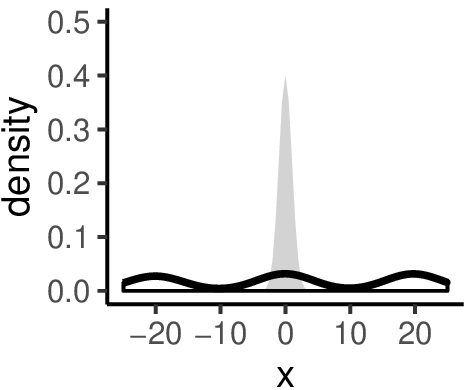}\\
\includegraphics[scale=0.7]{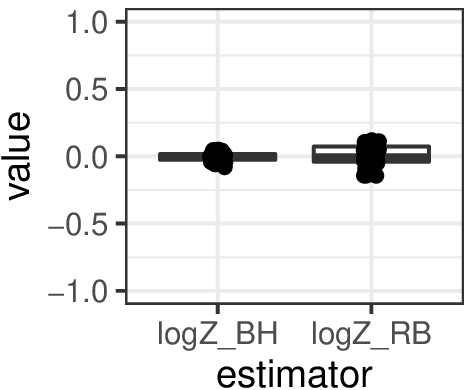}
\includegraphics[scale=0.7]{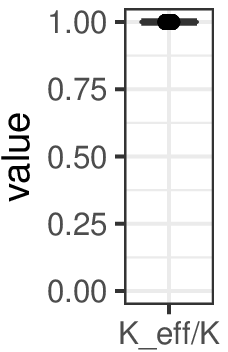}
\caption{$K=3; m=0.5; s=2$.}
\end{subfigure}
~
\begin{subfigure}[l]{0.48\linewidth}
\centering
\includegraphics[scale=0.7]{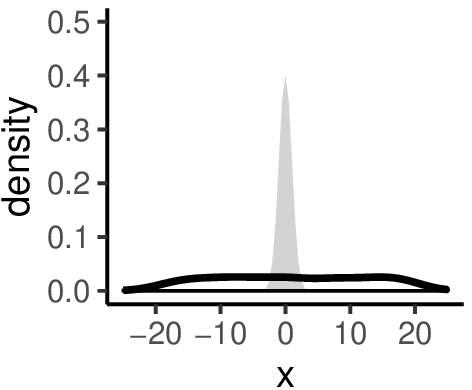}\\
\includegraphics[scale=0.7]{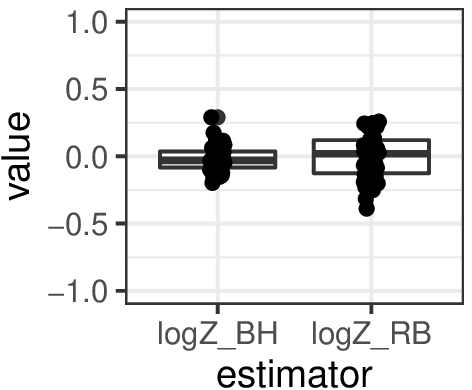}
\includegraphics[scale=0.7]{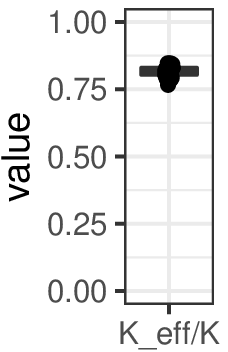}
\caption{$K=300; m=0.5; s=2$.}
\end{subfigure}
~
\begin{subfigure}[l]{0.48\linewidth}
\centering
\includegraphics[scale=0.7]{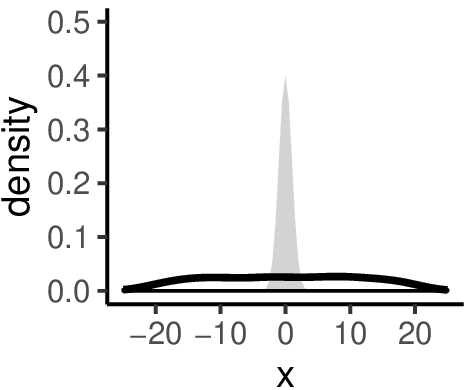}\\
\includegraphics[scale=0.7]{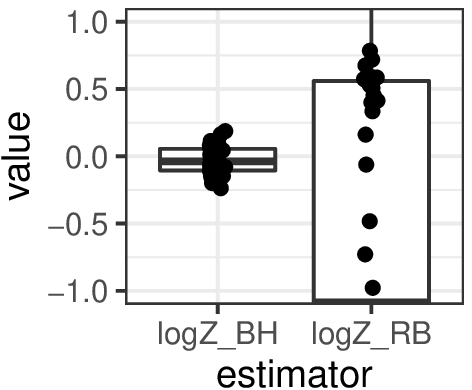}\includegraphics[scale=0.7]{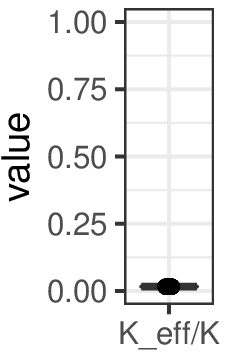}
\caption{$K=30,000; m=0.5; s=2$.}
\end{subfigure}
~
\begin{subfigure}[l]{0.48\linewidth}
\centering
\includegraphics[scale=0.7]{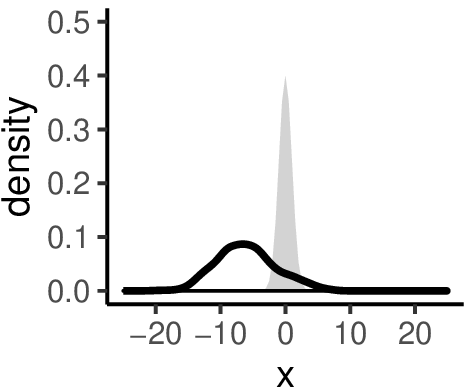}\\
\includegraphics[scale=0.7]{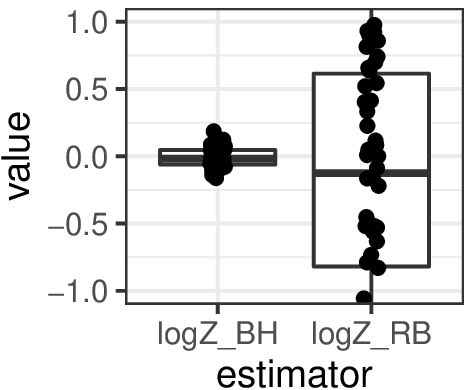}
\includegraphics[scale=0.7]{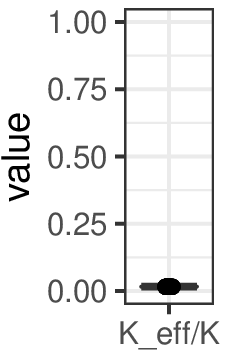}
\caption{$K=30,000; m=0.35; s=20$}
\end{subfigure}

}\caption{\label{fig:Fig1} Top: true target density (shaded area) and proposal density from
a sample $N=500$ (solid line). Bottom left: boxplots for $\log(\widehat{Z}_{BH})$ and $\log(\widehat{Z}_{RB})$ balance heuristic and Rao-Blackwellized for different values of $K$, $m$, $s$, and
equivalent computational costs. Bottom right: boxplots for proportion
$K_{eff}/K$.}
\end{figure}

\end{example}
We now explore further the balance heuristic estimator, noting that it can be expressed in terms of distributions on an extended space. This alternative representation will prove useful for reducing
the variance of the estimator by means of an annealing procedure.

\subsection{Extended-space representation}

The variables involved in the
computation of \eqref{eq:Z_BH} are $\mathcal{S}_{1}=\left\{ \left\{ X_{ij}\right\} _{j=1}^{N_{i}},N_{i}\right\} _{i=1}^{K}$,
where $\left\{ X_{ij}\right\} _{j=1}^{N_{i}}\stackrel{iid}{\sim}q_{i}\left(\cdot\right)$
for each $i\in\left\llbracket 1,K\right\rrbracket $ and, as discussed
before, the vector $N_{1:K}$ can be drawn from any appropriate distribution
satisfying $\sum_{i=1}^{N}N_{i}=N$. The alternative expression for $\widehat{Z}_{BH}$
in \eqref{eq:Z_BH_alt} involves the set of variables $\mathcal{S}_{2}=\left\{ \left(X_{n},L_{n}\right)\right\} _{n=1}^{N}$,
where $X_{n}\mid L_{n}\sim q_{L_{n}}\text{\ensuremath{\left(\cdot\right)}}$
and $L_{n}\sim\alpha\left(\cdot\right)$ for each $n\in\left\llbracket 1,N\right\rrbracket $.
The sets of variables $\mathcal{S}_{1}$ and $\mathcal{S}_{2}$ are
equivalent assuming in the former that $N_{1:K}\sim Mult\left(N,\alpha_{1:K}\right)$
and taking $\alpha_{k}=\alpha\left(k\right)$ for each $k\in\left\llbracket 1,K\right\rrbracket $;
therefore the variables in $\mathcal{S}_{1}$ can be expressed in
terms of those in $\mathcal{S}_{2}$ (and vice-versa) as follows
\begin{align}
N_{i} & =\sum_{n=1}^{N}\mathds{1}\left(L_{n}=i\right),\quad X_{ij} =X_{T_{ij}}, \label{eq:altRep}\\
&\text{where }T_{ij}=\inf\left\{ n\geq1\mid\sum_{m=1}^{n}\mathds{1}\left(L_{m}=i\right)=j\right\} .\nonumber 
\end{align}
The variable $T_{ij}$ represents the subindex in the sample
$\mathcal{S}_{2}$ for which the label $L_{n}=i$ for the $j$th
time.

The extended-space target, which results in the balance heuristic estimator in \eqref{eq:Z_BH_alt},
is
\begin{align*}
\eta_{BH}\left(x_{1:N},l_{1:N}\right) &  =\sum_{n=1}^{N}\left\{ \frac{\pi\left(x_{n}\right)q_{l_{n}}\left(x_{n}\right)\alpha\left(l_{n}\right)}{\sum_{m=1}^{N}q_{l_{m}}\left(x_{n}\right)}\prod_{m\neq n}\left\{ q_{l_{m}}\left(x_{m}\right)\alpha\left(l_{m}\right)\right\} \right\} ,
\end{align*}
noting that it is obtained by multiplying the estimator $\widehat{Z}_{BH}$
with the joint proposal density 
\[
\bar{q}^{\otimes N}\left(x_{1:N},l_{1:N}\right)=\prod_{n=1}^{N} \left\{ q_{l_{n}}\left(x_{n}\right)\alpha\left(l_{n}\right) \right\}
\]
 of the variables in $\mathcal{S}_{2}$ and dividing by $Z$ for normalisation
purposes. Therefore, the balance heuristic estimator is the same as an importance sampling with target
$\eta_{BH}$ using a single importance point. Furthermore, we can
extend the target beyond this since the sum in the expression above
can be seen as the marginalisation of a discrete variable $n$ on
the set $\left\llbracket 1,N\right\rrbracket $; doing this leads
to the following extended target on the space $\left\llbracket 1,N\right\rrbracket \times\mathcal{X}^{N}\times\left\llbracket 1,K\right\rrbracket ^{N}$
\begin{align}
\eta_{BH}\left(n,x_{1:N},l_{1:N}\right) & =\pi\left(x_{n}\right)\frac{q_{l_{n}}\left(x_{n}\right)}{\sum_{m=1}^{N}q_{l_{m}}\left(x_{n}\right)}\alpha\left(l_{n}\right)\prod_{m\neq n}\left\{ q_{l_{m}}\left(x_{m}\right)\alpha\left(l_{m}\right)\right\} .\label{eq:ext_taget_BH}
\end{align}
\begin{remark}
The target $\pi$ cannot be recovered by marginalising or
conditioning any of the variables from $\eta_{BH}$; nevertheless,
by construction the normalising constant for $\eta_{BH}$ is still
$Z$.
\end{remark}

An interesting consequence from the previous construction is the expression for the
full conditional of $X_{n}$, which is different to $\pi$ and given
by
\begin{equation}
\eta_{BH}\left(x_{n}\mid n,x_{-n},l_{1:N}\right)=\pi\left(x_{n}\right)\frac{q_{l_{n}}\left(x_{n}\right)}{\sum_{m=1}^{N}q_{l_{m}}\left(x_{n}\right)}/E_{X\sim\pi}\left(\frac{q_{l_{n}}\left(X\right)}{\sum_{m=1}^{N}q_{l_{m}}\left(X\right)}\right).\label{eq:ext_target_fullCond}
\end{equation}
In \eqref{eq:ext_target_fullCond}, $\pi$ is multiplied by a uniformly bounded
weight that assesses the suitability of $x_{n}$ to the corresponding
conditional $q_{l_{n}}$ versus the other sampled conditionals $q_{l_{m}}$,
this will be important in the next section when trying to mimic balance heuristic in the intractable scenario.

Working on artificially-extended spaces can shed light to difficult
problems, examples include the pseudo-marginal algorithm \citep{Beaumont_2003,AndrieuNRoberts_2009}
for dealing with intractable posteriors, the pseudo-extended
method from \citet{NemethETal_2017} for efficient exploration
of complex posteriors, or the tempered Gibbs sampler in \citet{ZanellaNRoberts_2018}
useful in high-dimensions. In the case of balance heuristic,
the extended representation in \eqref{eq:ext_taget_BH} will prove useful when introducing a sequence of intermediate targets for reducing the variance of the estimator. This idea goes back
to \citet{Jarzynski_1997,Neal_2001} and is commonly known as annealed importance sampling, which can be described as
bridging a proposal with the desired target. Popular annealing schemes include
arithmetic and geometric (see e.g. \citealp{KaragNAndrieu_2013} in
the context of reversible-jump MCMC). We restrict to purely-geometric
and semi-geometric schemes, which are now described for the balance heuristic extended distribution.

Consider a strictly increasing sequence $\left\{ \gamma_{t}\right\} _{t=0}^{T}$
such that $\gamma_{0}=0$ and $\gamma_{T}=1$. The density of the
$t$th intermediate distribution in a purely-geometric scheme
is
\begin{align*}
\eta_{BH,t}^{(pg)}\left(n,x_{1:N},l_{1:N}\right) & \propto\left[\eta_{BH}\left(n,x_{1:N},l_{1:N}\right)\right]^{\gamma_{t}}\left[\bar{q}^{\otimes N}\left(x_{1:N},l_{1:N}\right)\right]^{1-\gamma_{t}}\\
 & \propto\left(\frac{\tilde{\pi}\left(x_{n}\right)}{\sum_{m=1}^{N}q_{l_{m}}\left(x_{n}\right)}\right)^{\gamma_{t}}\bar{q}^{\otimes N}\left(x_{1:N},l_{1:N}\right) =\tilde{\eta}_{BH,t}^{(pg)}\left(n,x_{1:N},l_{1:N}\right).
\end{align*}
An alternative is to consider a slight modification, which we have
termed the semi-geometric scheme,
\begin{align*}
\eta_{BH,t}^{(sg)}\left(n,x_{1:N},l_{1:N}\right) & \propto\frac{\tilde{\pi}\left(x_{n}\right)^{\gamma_{t}}}{\sum_{m=1}^{N}q_{l_{m}}\left(x_{n}\right)^{\gamma_{t}}}\bar{q}^{\otimes N}\left(x_{1:N},l_{1:N}\right) =\tilde{\eta}_{BH,t}^{(sg)}\left(n,x_{1:N},l_{1:N}\right).
\end{align*}
In either case, the intermediate distributions are well defined since
the unnormalized densities $\tilde{\eta}_{BH,t}^{(pg)}$ and $\tilde{\eta}_{BH,t}^{(sg)}$
are integrable on the space $\left\llbracket 1,N\right\rrbracket \times\mathcal{X}^{N}\times\left\llbracket 1,K\right\rrbracket ^{N}$,
noting that for either scheme
\[
\eta_{BH,0}^{(\cdot)}=\frac{1}{N}\bar{q}^{\otimes N}\qquad\text{and}\qquad\eta_{BH,T}^{(\cdot)}=\eta_{BH}.
\]

With the aforementioned sequence of distributions, one can implement a standard annealed importance sampling procedure. Algorithm \ref{alg:AIS} describes this procedure,
where for simplicity we have only retained the subscript $t$ in  $\tilde{\eta}^{(\cdot)}_{BH,t}$ since the algorithm remains
valid for any choice of intermediate targets. From
the annealed importance sampling literature we know that for any $T\geq1$ the resulting estimate
$\widehat{Z}_{T,AIS}$ is also unbiased and $\mbox{var} \left( \widehat{Z}_{T,AIS}\right) \leq \mbox{var} \left( \widehat{Z}_{BH}\right)$.

\begin{algorithm}[!ht]
\caption{Standard annealed importance sampling:}\label{alg:AIS}
Draw $\theta_{0}=\left(X_{1:N},L_{1:N}\right)\sim\bar{q}^{\otimes N}\left(\cdot\right)$\;
Compute $ \tilde{W}_{1}\left(\theta_{0}\right) =\nicefrac{\sum_{n=1}^{N}\tilde{\eta}_{1}\left(n,\theta_{0}\right)}{\bar{q}^{\otimes N}\left(\theta_{0}\right)}$\;
\For{$t\in\left\llbracket 1,T-1\right\rrbracket $}{
\For{$t\in\left\llbracket 1,T-1\right\rrbracket $}{
Simulate $\theta_{t} \sim\mathcal{K}_{t}\left(\cdot\mid\theta_{t-1}\right)$, where $\mathcal{K}_{t}$ is an $\eta_{BH,t}$-invariant probability kernel\;
Compute $\tilde{W}_{t+1}\left(\theta_{t}\right) =\nicefrac{\sum_{n=1}^{N}\tilde{\eta}_{t+1}\left(n,\theta_{t}\right)}{\sum_{n=1}^{N}\tilde{\eta}_{t}\left(n,\theta_{t}\right)}$\;}
}
Output $\widehat{Z}_{T,AIS} =\prod_{t=1}^{T}\tilde{W}_{t}.$
\end{algorithm}

The weights $\left\{ \tilde{W}_{t}\right\} _{t=1}^{T}$
are computed marginalising over the artificial discrete variable $n$
in each intermediate target. It is possible to choose $\mathcal{K}_{t}$
as the resulting kernel from an MCMC algorithm targeting the extended target $\eta_{BH,t}$,
noting that the perturbation of the labels $l_{1:N}$ is not strictly
needed. A much simpler choice, since devising
an efficient MCMC targeting $\eta_{BH,t}$ is not straightforward,
is to implement a collapsed Gibbs sampler targeting the conditional
$\eta_{BH,t}\left(\cdot\mid l_{1:N}\right)$, in which one alternates
sampling $n^{*}\sim\eta_{BH,t}\left(\cdot\mid x_{1:N,}l_{1:N}\right)$
and $X_{-n^{*}}^{*}\sim\eta_{BH,t}\left(\cdot\mid n^{*},x_{n^{*}},l_{1:N}\right)$
which are in fact distributions easy to sample from. Performing this
Gibbs-type move can be more efficient than an MCMC-type move on the
space $\left\llbracket 1,N\right\rrbracket \times\mathcal{X}^{N}$,
as noted later in some examples. We stress that the previous
annealed importance sampling process is implemented using a single importance point $\theta_{0}=\left(X_{1:N},L_{1:N}\right)$,
consisting of a system of $N$ particles, that evolves from $t=0$
to $t=T-1$.

Performing a standard annealed importance sampling procedure comes with the extra
cost of computing the weights $\left\{ \tilde{W}_{t}\right\} _{t}$.
These involve a ratio of
two sums, each consisting of $N$ terms, leading to a total cost of $\mathcal{O}\left( TN^2 \right)$.
A different approach, that delays the marginalisation of the variable
$n$ until the end, is presented in Algorithm \ref{alg:mod_AIS} which
we have termed modified annealed importance sampling. In contrast to the standard procedure
in Algorithm \ref{alg:AIS}, the modified version is embarrassingly parallel since the perturbation and re-weighting steps can be performed independently for
each $x_{n}$, provided $l_{1:N}$ is fixed. This is still a process with overall cost $\mathcal{O}\left( TN^2 \right)$, but it can be computed much faster using $N$ simultaneous processes, each with cost $\mathcal{O}\left( TN \right)$.
For either scheme (purely-geometric or semi-geometric), the weight $\tilde{W}_{t}^{(n)}$ only depends on the variables
$X_{n}$ and $L_{1:N}$ within $\theta_{t-1}$; in order to keep the
notation simple we write $\theta_{t}\left[X_{n},L_{1:n}\right]$ to
denote this case. Similarly, the kernel $\mathcal{K}_{t,n}$ only
updates the variable $X_{n}$ within $\theta_{t}$, denoted as $\theta_{t}\left[X_{n}\right]$,
and is chosen to be invariant with respect to the conditional distribution
$\eta_{BH,t}\left(dx_{n}\mid n,l_{1:N}\right)$.
A simple choice for $\mathcal{K}_{t,n}$ is the resulting kernel from
an MCMC move on $\mathcal{X}$, which is potentially easier to tune
up than that from $\mathcal{K}_{t}$ in Algorithm \ref{alg:AIS}.

\begin{algorithm}[!ht]
\caption{Modified annealed importance sampling:}\label{alg:mod_AIS}
Draw $\theta_{0}=\left(X_{1:N},L_{1:N}\right)\sim\bar{q}^{\otimes N}\left(\cdot\right)$ \;
\For{$n\in\left\llbracket 1,N\right\rrbracket$}{
Compute $\tilde{W}_{1}^{\left(n\right)} \left(\theta_{0}\left[x_{n},l_{1:N}\right]\right) =\nicefrac{\tilde{\eta}_{1}\left(n,\theta_{0}\right)}{\bar{q}^{\otimes N}\left(\theta_{0}\right)}$ \;
\For{$t\in\left\llbracket 1,T-1\right\rrbracket $}{
Simulate $\theta_{t}\left[X_{n}\right] \sim\mathcal{K}_{t,n},$ where $\mathcal{K}_{t,n}$ is an $\eta_{t}\left(dx_{n}\mid n,l_{1:N}\right)$-invariant probability kernel\;
Compute $\tilde{W}_{t+1}^{\left(n\right)} \left(\theta_{t}\left[x_{n},l_{1:N}\right]\right) =\nicefrac{\tilde{\eta}_{t+1}\left(n,\theta_{t}\right)}{\tilde{\eta}_{t}\left(n,\theta_{t}\right)}$\;}
}
Output $\widehat{Z}_{T,mAIS} =\sum_{n=1}^{N}\prod_{t=1}^{T}\tilde{W}_{t}^{\left(n\right)}$.
\end{algorithm}

The following result shows that Algorithm \ref{alg:mod_AIS} produces
an unbiased estimator for $Z$, its proof can be found in Appendix 1.
\begin{theorem}
\label{thm:modAIS}The estimator $\widehat{Z}_{T,mAIS}$ 
from Algorithm \ref{alg:mod_AIS} satisifes $E\left(\widehat{Z}_{T,mAIS} \right)=Z$.
\end{theorem}
A sequential Monte Carlo process
(see \citealp{DoucetETal_2001,MoralETal_2006}) could be constructed
inspired by the previous algorithms, e.g. one would require a
set of $M$ initial particles $\left\{ \theta_{0}^{(i)}\right\} _{i=1}^{M}$
that are propagated in a similar fashion as in Algorithm \ref{alg:mod_AIS} with
the possibility of introducing resampling between particles.
In this respect, sequential Monte Carlo using multiple proposals has been explored in
\citet{LiETal_2016}. Additionally, it is not obvious how one would
perform resampling within Algorithm \ref{alg:mod_AIS} that could potentially retain \textit{good} values of $X_{1:N}$ without breaking the unbiasedness.

\begin{example}[Running example continued]

Figure \ref{fig:Fig3} compares the estimators $\widehat{Z}_{T,AIS}$ and $\widehat{Z}_{T,mAIS}$ for different specifications of the beta-binomial distribution. The MCMC kernels in standard annealed importance sampling and the modified annealed importance sampling are different
in nature since in the former case the moves are performed on the
space $\mathcal{X}^{N}$, whereas for the latter (as described in Algorithm
\ref{alg:mod_AIS}) we perform $N$ MCMC moves on $\mathcal{X}$.
In both cases we considered 10 Metropolis--Hastings iterations with
Gaussian proposals, on their respective dimension.
In addition, the number of intermediate distributions considered were
20. We present only the purely-geometric case since the semi-geometric does not differ substantially.
Extra plots contained in the supplementary material compare the two schemes.

Subfigure (a) corresponds to the same scenario as Subfigure (c) in
Fig. \ref{fig:Fig1}, observing that the standard and modified
annealed importance sampling processes lead to a reduction in variance. This is less apparent in the standard annealed importance sampling with MCMC moves, where our
simple MCMC implementation proves inadequate. The proposal in Subfigure (b) is shifted from the origin as in Subfigure (d) from Fig. \ref{fig:Fig1}; here the Gibbs version for standard annealed importance sampling and the modified annealed importance sampling show the best performance. Subfigure (c) shows
the results of a more challenging scenario, in which the marginal
proposal for $X$ is very concentrated and none of the sampled values
lies near the target's core region. The Gibbs version shows an improvement over the MCMC approach but the bias is still
present. This is because for perturbing $X_{-n}$ we use $q_{l_{i}}\left(\cdot\right)$, for $i\in\left\llbracket 1,N\right\rrbracket \setminus\left\{ n\right\} $,
which are unlikely to draw values near the target. Updating the values of $l_{1:N}$ at each step of the annealed importance sampling
would be beneficial, but with the complication of sampling from the
full conditional for $L_{1:N}\mid n,X_{1:N}$ on the space $\left\llbracket 1,K\right\rrbracket ^{N}$.
In contrast, the modified annealed importance sampling estimates shows the best improvements.
\begin{figure}[!ht]
\centering
\begin{subfigure}{\linewidth}
\centering
\includegraphics[scale=0.7]{density_K30000_m0\lyxdot 5_s2}\includegraphics[scale=0.7]{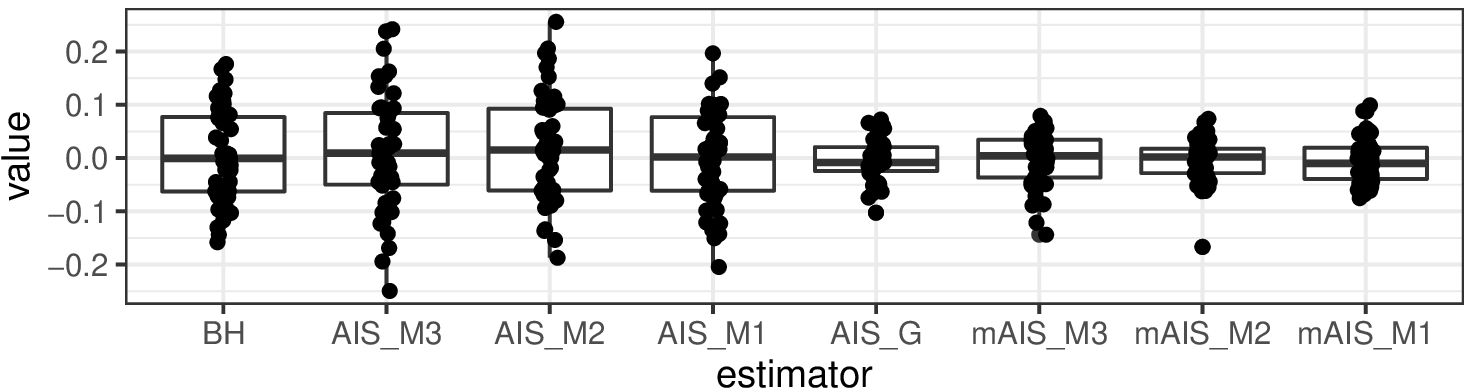}
\caption{$m=0.5, s=2$.}
\end{subfigure}

\begin{subfigure}{\linewidth}
\centering
\includegraphics[scale=0.7]{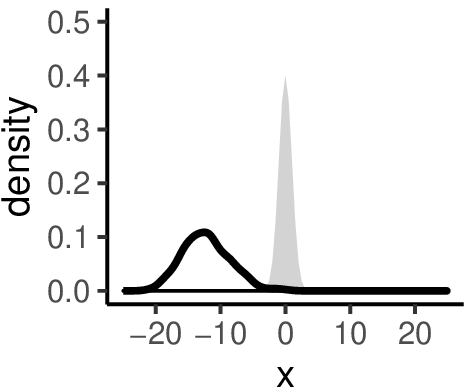}\includegraphics[scale=0.7]{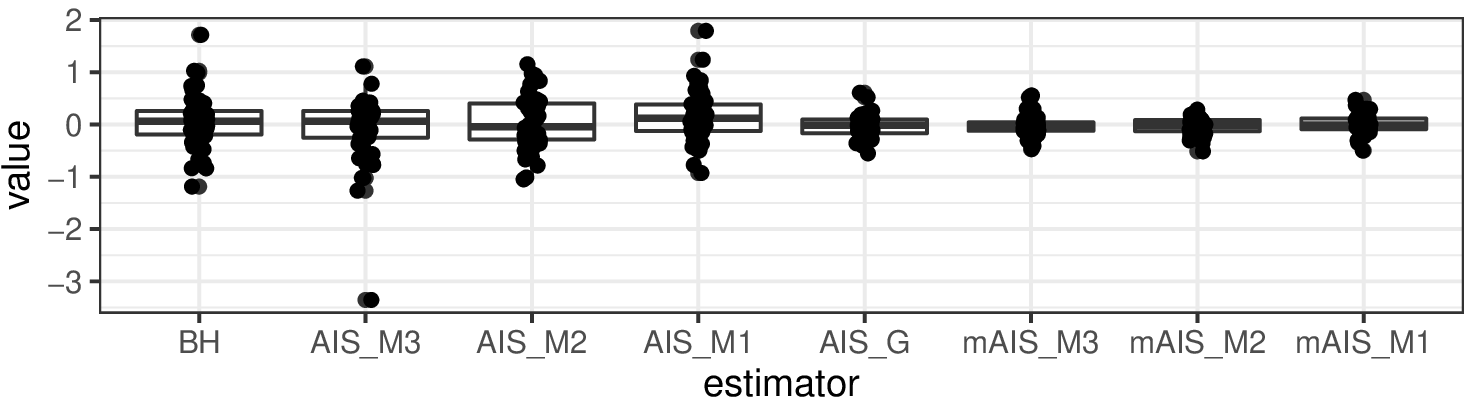}
\caption{$m=0.2, s=20$.}
\end{subfigure}

\begin{subfigure}{\linewidth}
\centering
\includegraphics[scale=0.7]{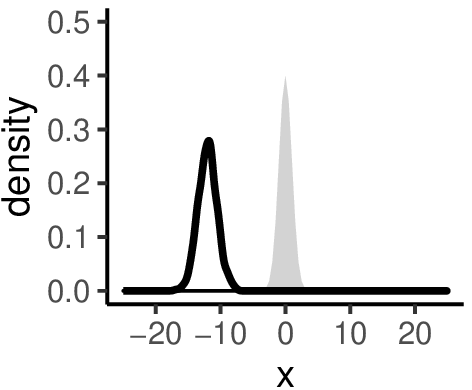}\includegraphics[scale=0.7]{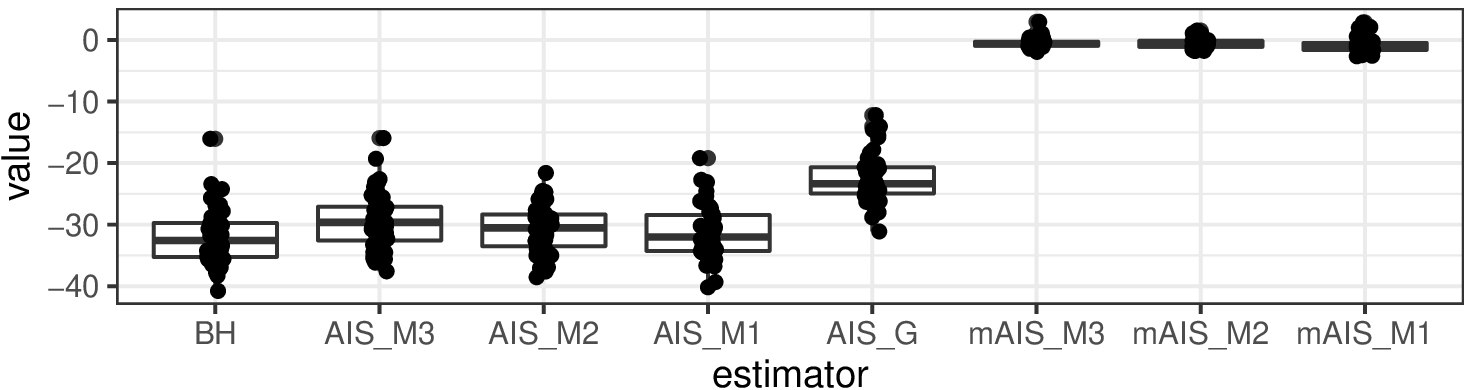}
\caption{$m=0.2, s=\infty$.}
\end{subfigure}

\caption{\label{fig:Fig3}Left: true target density (shaded area) and proposal density from
a sample $N=500$ (solid line). Right: boxplots comparing normalising constant estimates in log-scale for different values of $m$ and $s$ and different MCMC moves. All cases consider $K=30,000$, a purely-geometric scheme and $T=21$.}
\end{figure}
\end{example}

\section{Estimators for intractable proposals\label{sec:intractable_proposals}}
\subsection{Motivation}

We now explore the case when the joint proposal for $\left(X,L\right)$
(denoted by $\bar{q}$) shows some level of intractability, in the
sense that the factorisation into the product of the conditional density
for $X\mid L$ and the marginal density for $L$ is not analytically
available. This contrasts with the scenario discussed in the previous
section, where by construction the joint proposal density for $\left(X,L\right)$
is factorized as $\bar{q}\left(x,l\right)=q_{l}\left(x\right)\alpha\left(l\right)$,
with $q_{l}$ and $\alpha$ known. The setting considered in this
section commonly arises when an artificial label is introduced for
imposing some order on the variables of interest. or example, in multi-object
tracking the aim is to infer the present states or paths of multiple
moving objects and, in order to avoid ambiguity, the objects need
to be labelled according to some variable of interest (this could
be a time or spatial component), see e.g. \citet{JiangETal_2015,LiETal_2017,JiangETSingh_2018}.
This reasoning also applies to mixture models and clustering, for
which a labelling of components may be introduced in order to avoid
identifiability problems with the parameters of interest. We have
noted some of these aspects in \citet{EverittETal_2016}, where ordering
of components leads to the aforementioned intractability of the joint
proposal $\bar{q}$.

In essence, whenever we deal with ordering there is the potential
problem of dealing with an intractable proposal. We develop this
idea further through a conceptual example found in Appendix 2. To simplify, the intractable setting we consider throughout this section
can be stated as follows: joint draws $\left(X,L\right)\sim\bar{q}\left(\cdot\right)$
are available, $L$ is a discrete variable on the finite support $\left\llbracket 1,K\right\rrbracket $,
we want to avoid computing the marginal density for $X$ since it
is expensive, and the conditional $q_l\left(x\right)$ is not
analytically available.

\subsection{Combining multiple estimators}

Whilst in theory $\bar{q}$ can always be decomposed as $\bar{q}\left(x,l\right)=q_{l}\left(x\right)\alpha\left(l\right)$,
in this case we assume that expressions for $\alpha$ and $q_{l}$
are not at hand. Then, how can one use $\bar{q}$ to estimate the
normalising $Z$ from the target $\pi$? One approach is to consider
the following estimator
\begin{align}
\widehat{Z}_{\beta} & =\frac{\tilde{\pi}\left(X\right)\beta_{X}\left(L\right)}{\bar{q}\left(X,L\right)},\qquad \left(X,L\right)\sim\bar{q}\left(\cdot\right),\label{eq:Z_joint}
\end{align}
for some arbitrary auxiliary distribution $\beta_{x}$ on $\left(\left\llbracket 1,K\right\rrbracket ,\mathcal{P}\left(\left\llbracket 1,K\right\rrbracket \right)\right)$,
that may depend on the value of $X$. The optimal choice for $\beta_{x}$
(the one minimising the variance of \eqref{eq:Z_joint}) is
\begin{align*}
\beta_{x}^{opt}\left(l\right) & =\frac{\bar{q}\left(x,l\right)}{\sum_{k=1}^{K}\bar{q}\left(x,k\right)},
\end{align*}
which is in fact the Rao-Blackwellized estimator in \eqref{eq:Z_opt} from the previous
section. Therefore, it is optimal in terms of variance
within the family of estimators $\left\{ Z_{\beta}\right\} _{\beta}$.
However, when the number of labels $K$ is very large
then computing the denominator $\beta_{x}^{opt}$ will be expensive
or impractical. The simplest approach we can think of is to choose
$\beta^{unif}\left(l\right)=K^{-1}$, i.e. a uniform distribution
for the labels; unfortunately, this could result in an estimator with very
large variance depending on the mismatch between the unavailable distribution
$\alpha$ and $\beta^{unif}$, as seen in the following example.
\begin{example}[Running example continued]
In Figure \ref{fig:Fig4} we compare estimates using $\beta^{unif}$ and $\beta_{x}^{opt}$. Subfigure (a) consider $K=3$ labels,
observing that the estimator using $\beta^{unif}$
produces values around $\log(K)$. This occurs due to $\tilde{\pi}\left(x\right)\approx0$ whenever the associated
$l$ for such $x$ is either 1 or 3, this means that we are essentially
failing to explore 2/3 of the extended target $\pi\left(x\right)\beta^{unif}\left(l\right)$,
which introduces the apparent bias. Theoretically
the estimator converges to the actual value of $Z$, however,
in practical terms we observe a bias in the estimates due to the high
variance associated to the estimator.
One could correct this undesirable behaviour
by discarding samples that are located far from high posterior regions,
or by modifying the auxiliary distribution $\beta_{x}$ accordingly.
In more complex situations, e.g as in Subfigure (b) where $K=300$,  implementing such strategies is
not straightforward. In contrast, the estimates produced
by $\beta_{x}^{opt}$ appear to be less variable since they are centred
around the truth, with the caveat that they are expensive to compute
if $K$ becomes very large.
\begin{figure}[!ht]
\centering
\begin{subfigure}{0.48\linewidth}
\centering
\includegraphics[scale=0.7]{{density_K3_m0\lyxdot 5_s2}.eps}\includegraphics[scale=0.7]{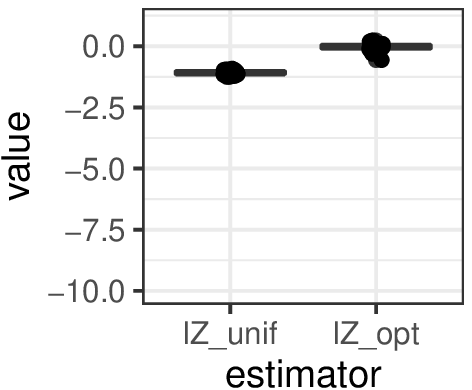}
\caption{$K=3$.}
\end{subfigure}
~
\begin{subfigure}{0.48\linewidth}
\centering
\includegraphics[scale=0.7]{{density_K300_m0\lyxdot 5_s2}.eps}\includegraphics[scale=0.7]{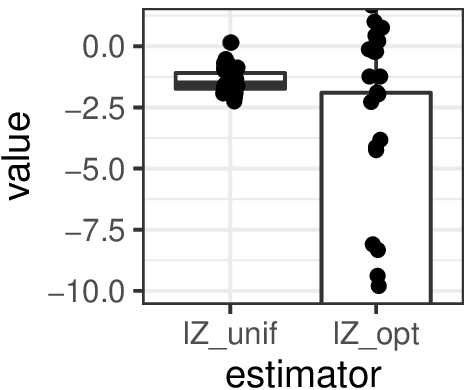}
\caption{$K=300$.}
\end{subfigure}

\caption{\label{fig:Fig4}Left: true target density (shaded area) and proposal density from
a sample $N=500$ (solid line). Right: boxplots for resulting estimates using $\beta^{unif}$ (left boxplot) and $\beta_{x}^{opt}$ for different values of $m$ and equivalent computational costs. All cases consider $m=0.5$
and $s=2$.}
\end{figure}
\end{example}
Potentially, a better possibility for $\beta_{x}$ would be to consider $\beta_{x}\left(l\right)=\alpha\left(l\right)$,
but an expression for the density $\alpha$ is not available.
Nevertheless, we can obtain an unbiased estimator through a sample
$\left\{ \left(X_{n},L_{n}\right)\right\} _{n=1}^{N}\stackrel{iid}{\sim}\bar{q}\left(\cdot\right)$,
and using the relationship in \eqref{eq:altRep} between the variables
$\left\{ \left(X_{n},L_{n}\right)\right\} _{n=1}^{N}$ and $\left\{ \left\{ X_{ij}\right\} _{j=1}^{N_{i}},N_{i}\right\} _{i=1}^{K}$,
implying $N_{1:K}\sim Mult\left(\cdot\mid N,\alpha_{1:K}\right)$.
Using standard results from the multinomial distribution we obtain
the following result, which is  proved in Appendix 1.
\begin{lemma}
\label{lem:multEstims}Consider a sample $\left\{ \left(X_{i},L_{i}\right)\right\} _{i=1}^{N}\stackrel{iid}{\sim}\bar{q}\left(\cdot\right)$,
using \eqref{eq:altRep} define for $i\in\text{\ensuremath{\left\llbracket 1,K\right\rrbracket }}$
\begin{align}
\widehat{Z}_{i} & =\frac{1}{N}\sum_{j=1}^{N_{i}}\frac{\tilde{\pi}\left(X_{ij}\right)}{\bar{q}\left(X_{ij},i\right)}.\label{eq:Zi_comb}
\end{align}
Then, for each $i\in\text{\ensuremath{\left\llbracket 1,K\right\rrbracket }}$
\begin{align*}
E\left(\widehat{Z}_{i}\right) & =Z,\quad \mbox{var} \left(\widehat{Z}_{i}/Z\right) =\frac{1}{N^{2}}\left( E_{X\sim\pi}\left( \frac{\pi\left(X\right)}{\bar{q}\left(X,i\right)}\right)-1\right);
\end{align*}
and for any pair $\left(i,j\right)\in\left\llbracket 1,K\right\rrbracket ^{2}$
such that $i\neq j$
\begin{align*}
\mbox{cov}\left(\widehat{Z}_{i}/Z,\widehat{Z}_{j}/Z\right) & =-\frac{1}{N}.
\end{align*}
\end{lemma}
By the previous result, any linear combination of the estimators
$\widehat{Z}_{1:K}$ will also be unbiased. Due to the
negative correlation between the estimators, one can then speculate
how to efficiently combine the estimators $\widehat{Z}_{1:K}$? One
must be cautious since linear combinations of estimators
can lead to variables with higher variance. We could, e.g., use the
effective sample size associated to each estimator to inform the weights in the linear
combination as done in \citet{GramacyETal_2010,NguyenETal_2015};
however, this could be problematic in our context since some of the
$\widehat{Z}_{i}$ may be computed using only a handful of points.
This is particularly true when $K$ is much larger than $N$ and in
turn will lead to inaccurate results. It is also worth acknowledging
the recent work in \citet{OwenNZhou_2019}, where the authors investigate
linear combinations of unbiased and uncorrelated estimators obtained
from an adaptive IS process. The following theoretical result provides
an optimal set of weights for combining $\widehat{Z}_{1:K}$, in the
sense that they minimise the variance of the combined estimator. Its
proof can be found in Appendix 1.
\begin{theorem}
\label{thm:optimalWeights}Consider the estimators $\widehat{Z}_{1:K}$ defined in \eqref{eq:Zi_comb} using a sample $\left\{ \left(X_{i},L_{i}\right)\right\} _{i=1}^{N}\stackrel{iid}{\sim}\bar{q}\left(\cdot\right)$. The variance of the estimator
\begin{align*}
\widehat{Z}_{comb} & =\sum_{i=1}^{K}\nu_{i}\widehat{Z}_{i},\qquad\text{where }0\leq\nu_{i}\leq\sum_{j=1}^{K}\nu_{j}=1\text{ for all }i\in\left[1:K\right],
\end{align*}
attains its global minimum when 
\begin{align*}
\nu_{i} & =\nu_{i}^{opt}=\frac{\mathbf{e}^T_{i,K}\Sigma_{K}^{-1}\mathbf{1}_{1:K}}{\mathbf{1}^T_{1:K}\Sigma_{K}^{-1}\mathbf{1}_{1:K}},
\end{align*}
where $\Sigma_{K}$ is the variance-covariance matrix of $\widehat{Z}_{1:K}$,
$\mathbf{e}_{i,K}$ is the vector of size $K$ with all entries equal
to 0 except for the $i$th one which is 1, and $\mathbf{1}_{1:K}=\sum_{i=1}^{K}\mathbf{e}_{i,K}$.
\end{theorem}
Despite the previous straightforward result, computing or accurately
estimating accurately $\Sigma_{K}^{-1}$ may prove challenging. The expressions for the variance and covariance terms involve
$Z^{2}$, however, this is not needed since we just need 
$\Sigma_{K}$ up to a proportional constant. Letting $\tau_{i}=E_{X\sim\pi}\left( \pi\left(X\right) / \bar{q}\left(X,i\right) \right)$
for each $i\in\left\llbracket 1,K\right\rrbracket $,
the variance-covariance matrix has the following form
\begin{align*}
\Sigma_{K} & \propto\left[\begin{array}{cccc}
\tau_{1}-1 & -1 & \cdots & -1\\
-1 & \tau_{2}-1 & \cdots & -1\\
\vdots & \vdots & \ddots & -1\\
-1 & -1 & \cdots & \tau_{K}-1
\end{array}\right] 
 =\underbrace{\left[\begin{array}{cccc}
\tau_{1} & 0 & \cdots & 0\\
0 & \tau_{2} & \cdots & 0\\
\vdots & \vdots & \ddots & 0\\
0 & 0 & \cdots & \tau_{K}
\end{array}\right]}_{T_{K}}-\mathbf{1}{}_{1:K}\mathbf{1}^T{}_{1:K}.
\end{align*}
Computing the inverse of $T_{K}$ is more straightforward, indeed
using the Sherman--Morrison formula for matrix inversion we have that
\begin{align*}
\Sigma_{K}^{-1} & \propto T_{K}^{-1}\left(1+\mathbf{1}^T{}_{1:K}T_{K}^{-1}\mathbf{1}{}_{1:K}\right)-\left(\mathbf{1}^T{}_{1:K}T_{K}^{-1}\right)^{T}\left(\mathbf{1}'{}_{1:K}T_{K}^{-1}\right).
\end{align*}
The main difficulty lies now in computing (or at least accurately
estimating) $\tau_{i}^{-1}$.

A simple estimate of $\tau_{i}=Z^{-1}\int\frac{\tilde{\pi}\left(x\right)}{\bar{q}\left(x,i\right)}\pi\left(x\right)\mu\left(dx\right)$
would be 
\begin{align*}
\widehat{\tau_{i}} & =\left(\frac{1}{N}\sum_{m=1}^{N}W_{m}\right)^{-1}\sum_{n=1}^{N}\frac{W_{n}}{\sum_{m=1}^{N}W_{m}}\times\frac{\tilde{\pi}\left(X_{n}\right)}{\bar{q}\left(X_{n},i\right)}\\
 & =\frac{N}{\left(\sum_{m=1}^{N}W_{m}\right)^{2}}\sum_{n=1}^{N}W_{n}\times\frac{\tilde{\pi}\left(X_{n}\right)}{\bar{q}\left(X_{n},i\right)},
\end{align*}
where $W_{n}=W\left(X_{n},L_{n}\right)=\tilde{\pi}\left(X_{n}\right)K^{-1}/\bar{q}\left(X_{n},L_{n}\right)$.
The above estimator uses the full sample for computing
each $\widehat{\tau_{i}}$, and involves the crude estimate for $Z$
using $\beta^{unif}\left(l\right)=K^{-1}$ discussed earlier. Even though this estimate of $Z$ is usually a bad
choice, the estimate for $\tau_{i}$ might
not necessarily be disastrous. The reasoning behind this idea is that
a self-normalising estimate for expectations of the form
\[
E_{X\sim\pi}\left( h\left(X\right)\right) =Z^{-1}\int h\left(X\right)\tilde{\pi}\left(x\right)\mu\left(dx\right)
\]
 usually performs better than estimating separately the normalising
constant and the integral part. We now show an example using the vector $\widehat{\tau}_{1:K}$ for approximating $T_K$.

\begin{example}[Running example continued]
Observe from Figure \ref{fig:Fig5} that, for equivalent computational
costs, the variability of $\widehat{Z}_{comb}$ is
smaller than that of $\widehat{Z}_{RB}$. This is more noticeable
when $K$ is large (when compared to $N$), however for such cases
the estimates of $\widehat{Z}_{comb}$ are not centred around the
truth $Z=1$; in fact, the way $\widehat{Z}_{comb}$ was constructed
does not guarantee the resulting estimator is in general unbiased.
Theoretically, consistency still holds, i.e. as $N\rightarrow\infty$
the estimator $\widehat{Z}_{comb}$ will converge with probability one to $Z$; however, from the example it appears that we must take $N$
at least equal to $K$ in order for the bias to be negligible. This
behaviour is also confirmed in extra figures in the supplementary material, where results for larger $N$ and a different set of $K's$ are shown. Nevertheless, for moderate value of $K$
this new estimate is very competitive with respect to $\widehat{Z}_{RB}$
and even to $\widehat{Z}_{BH}$, which is mainly presented for comparison.
Additionally, it seems that the estimator works better when the marginal
proposal for $X$ is concentrated (leading to $K_{eff}\ll K$) as
considered in Fig. \ref{fig:Fig5}. In the supplementary material we show plots when $s=2$ (i.e. the marginal for $X$ is spread), observing that even when $N\geq K$ the resulting bias is not small.
\begin{figure}[!ht]
\centering
\begin{subfigure}{\linewidth}
\centering
\includegraphics[scale=0.7]{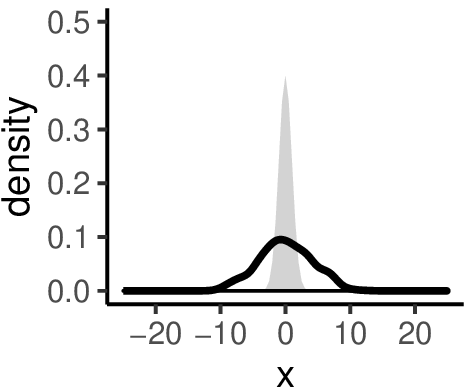}
\includegraphics[scale=0.7]{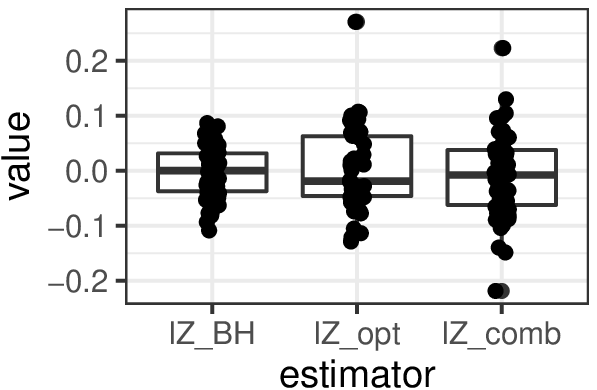}\includegraphics[scale=0.7]{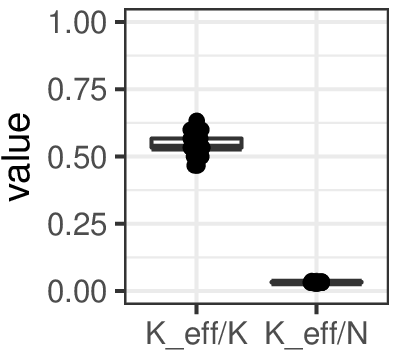}
\caption{ $K=30$.}
\end{subfigure}

\begin{subfigure}{\linewidth}
\centering
\includegraphics[scale=0.7]{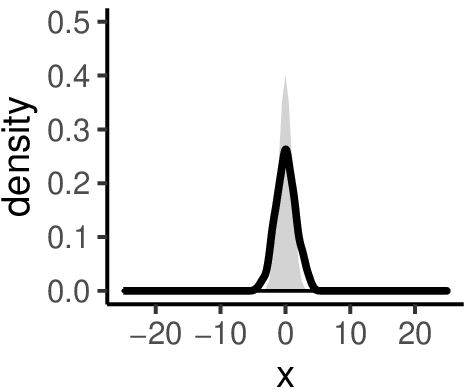}\includegraphics[scale=0.7]{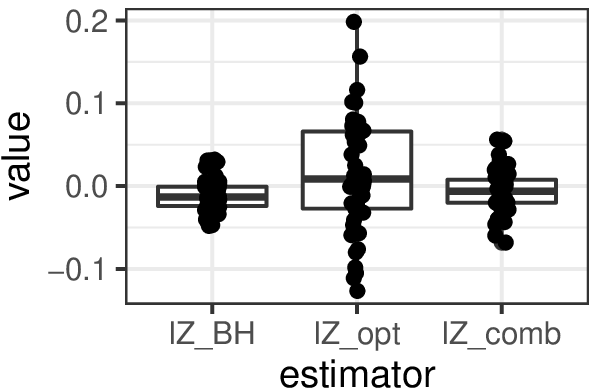}\includegraphics[scale=0.7]{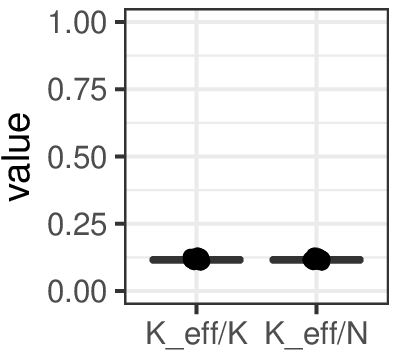}
\caption{ $K=500$.}
\end{subfigure}

\begin{subfigure}{\linewidth}
\centering
\includegraphics[scale=0.7]{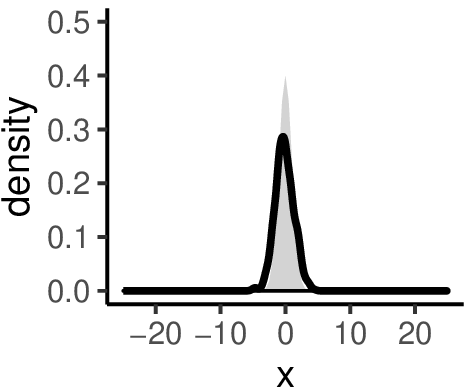}
\includegraphics[scale=0.7]{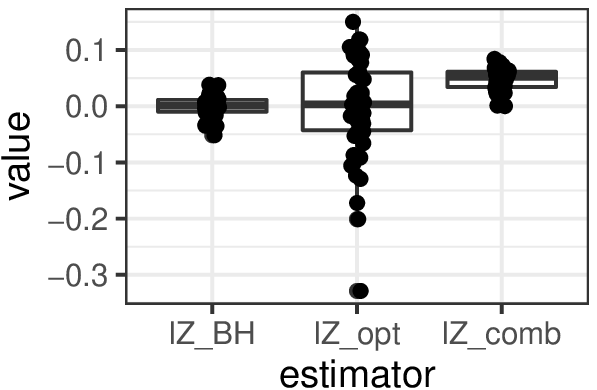}\includegraphics[scale=0.7]{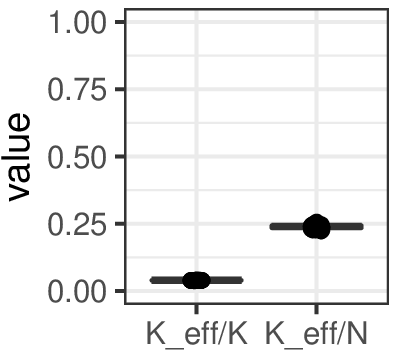}
\caption{ $K=3,000$.}
\end{subfigure}

\caption{\label{fig:Fig5}Left: true target density (shaded area) and proposal density from
a sample $N=500$ (solid line). Middle: boxplots comparing normalising constant estimates in log-scale for different values of $K$. Right: boxplots for proportions $K_{eff}/K$ and $K_{eff}/N$. All cases consider $m=0.5$ and $s=\infty$.}
\end{figure}
\end{example}
It turns out that $\widehat{Z}_{comb}$ can also be expressed in terms
of an extended target representation as in \eqref{eq:ext_taget_BH}.
Following similar steps for deriving $\eta_{BH}$, the extended target
for this case is
\begin{align}
\eta_{comb}\left(n,x_{1:N},l_{1:N}\right) & =\pi\left(x_{n}\right)N^{-1}\nu_{l_{n}}\prod_{m\neq n}\left\{ \bar{q}\left(x_{m},l_{m}\right)\right\} ,\label{eq:ext_target_comb}
\end{align}
and is valid for any vector $\nu_{1:K}$ such that $0\leq\nu_{i}\leq\sum_{j=1}^{K}\nu_{j}=1$
for every $i\in\left\llbracket 1,K\right\rrbracket $. Comparing $\eta_{BH}$
from \eqref{eq:ext_taget_BH} and $\eta_{comb}$ from \eqref{eq:ext_target_comb},
the balance heuristic target can be factorized in the
following way
\begin{align}
\eta_{BH} & \left(n,x_{1:N},l_{1:N}\right)=\eta_{BH}\left(x_{n}\mid n,x_{-n},l_{1:N}\right)\eta_{BH}\left(x_{-n}\mid n,l_{1:N}\right)\eta_{BH}\left(n,l_{1:N}\right)\nonumber \\
 & =\frac{\pi\left(x_{n}\right)\frac{q_{l_{n}}\left(x_{n}\right)}{\sum_{m=1}^{N}q_{l_{m}}\left(x_{n}\right)}}{E_{X\sim\pi}\left(\frac{q_{l_{n}}\left(X\right)}{\sum_{m=1}^{N}q_{l_{m}}\left(X\right)}\right)}\times\prod_{m\neq n}q_{l_{m}}\left(x_{m}\right)\times E_{X\sim\pi}\left( \frac{q_{l_{n}}\left(X\right)}{\sum_{m=1}^{N}q_{l_{m}}\left(X\right)}\right) \alpha^{\otimes N}\left(l_{1:N}\right),\label{eq:ext_target_BH_fact}
\end{align}
whereas
\begin{align*}
\eta_{comb}\left(n,x_{1:N},l_{1:N}\right) & =\eta_{comb}\left(x_{n}\mid n,x_{-n},l_{1:N}\right)\eta_{comb}\left(x_{-n}\mid n,l_{1:N}\right)\eta_{comb}\left(n,l_{1:N}\right)\\
 & =\pi\left(x_{n}\right)\times\prod_{m\neq n}q_{l_{m}}\left(x_{m}\right)\times N^{-1}\frac{\nu_{l_{n}}}{\alpha\left(l_{n}\right)}\alpha^{\otimes N}\left(l_{1:N}\right).
\end{align*}
The equations above show that the balance heuristic is more intricate than the combined estimators presented earlier. As commented in Section \ref{sec:MIS_BH},
the full conditional $\eta_{BH}\left(x_{n}\mid n,x_{-n},l_{1:N}\right)$
is different to $\pi$ which contrasts to $\eta_{comb}\left(x_{n}\mid n,x_{-n},l_{1:N}\right)$.
By incorporating a weight assessing the suitability of $x_{n}$ to
the associated conditional $q_{l_{n}}$ versus the other sampled conditionals
$q_{l_{m}}$, the balance heuristic estimator is able to produce more accurate estimates
of $Z$. A better approach in this intractable setting
would be one that resembles more to the balance heuristic.

\subsection{General framework: approximating balance heuristic}

In this section we explore approximations of balance heuristic when evaluations of
$q_{l}\left(x\right)$ and $\alpha\left(l\right)$ cannot be done
individually, i.e. only the joint density $\bar{q}\left(x,l\right)$
can be evaluated pointwise. Looking back at \eqref{eq:ext_target_BH_fact},
we can replace the conditionals $\left\{ q_{l_{n}}\right\} _{n=1}^{N}$ used in the first conditional density
with a set of positive functions $\Psi_{N}=\left\{ \psi_{n}\right\} _{n=\text{1}}^{N}$, which may depend on the sampled labels $l_{1:N}$; in addition, we also need to replace  $\alpha\left(l_{n}\right)$ with 
some approximation
$\varrho_{n}$, depending also possibly on $l_{1:N}$. The resulting extended target for this general framework is
\begin{align}
\eta_{GF}\left(n,x_{1:N},l_{1:N}\right)& \propto \pi\left(x_{n}\right)\frac{\psi_{n}\left(x_{n}\right)}{\sum_{m=1}^{N}\psi_{m}\left(x_{n}\right)}\varrho_{n}\prod_{m\neq n}\left\{ \bar{q}\left(x_{m},l_{m}\right)\right\} .\label{eq:eta_new}
\end{align}
In the previous equation there is a sign
of proportionality which has to do with the fact that, depending on
the choices for $\Psi_{N}$ and $\rho_{n}$, the expression may not be normalized. The resulting normalising
constant is given by the following result, and its proof can be found
in the Appendix 1.
\begin{proposition}
\label{prop:normConst_GF}The expression in \eqref{eq:eta_new} integrates
to
\begin{align*}
\mathfrak{Z}\equiv\mathfrak{Z}\left(\Psi_{N},\varrho_{n}\right) & =E_{L_{1:N}\sim\alpha^{\otimes N},X\sim\pi}\left( \frac{\sum_{n=1}^{N}\psi_{n}\left(X\right)\frac{\varrho_{n}}{\alpha\left(L_{n}\right)}}{\sum_{m=1}^{N}\psi_{m}\left(X\right)}\right).
\end{align*}
In particular, if $\psi\equiv 1$ then $\mathfrak{Z}=1$.
\end{proposition}
\begin{remark}
Not only the balance heuristic belongs to this framework, in which $\psi_{n}\left(x\right)=q_{l_{n}}\left(x\right)$
and $\varrho_{n}=\alpha\left(l_{n}\right)$, but also the combined
estimators method from the previous section does, when $\varrho_{n}=\nu_{l_{n}}$
and $\psi_{n}\equiv1$. 
\end{remark}
The extended target $\eta_{GF}$ leads to
the following estimator
\begin{align*}
\widehat{Z}_{GF} & =Z\mathfrak{Z}\sum_{n=1}^{N}\frac{\eta_{gf}\left(n,x_{1:N},l_{1:N}\right)}{\prod_{m=1}^{N}\left\{ \bar{q}\left(x_{m},l_{m}\right)\right\} } =\sum_{n=1}^{N}\frac{\tilde{\pi}\left(x_{n}\right)\psi_{n}\left(x_{n}\right)\varrho_{n}\left(l_{n}\right)}{\bar{q}\left(x_{n},l_{n}\right)\sum_{m=1}^{N}\psi_{m}\left(x_{n}\right)},
\end{align*}
which may prove useful in the intractable proposal scenario despite
the possibility of being biased with respect to $Z$. We now look
at two specific implementations of this idea.
\begin{itemize}
\item Consider $\psi_{n}\left(x\right)\equiv1$ and $\varrho_{n}=\left(K^{-1}+\sum_{m=1,m\neq n}^{N}\mathds{1}\left(l_{m}=l_{n}\right)\right)/N$
for all $n\in\left\llbracket 1,N\right\rrbracket $. Observe that
in this case $\varrho_{n}=\left(K^{-1}-1+N_{l_{n}}\right)/N$, hence
$\widehat{Z}_{GF}$ reduces to 
\begin{align*}
\widehat{Z}_{GF_{1}} & =\frac{1}{N}\sum_{n=1}^{N}\frac{\tilde{\pi}\left(x_{n}\right)}{\bar{q}\left(x_{n},l_{n}\right)}\left(\frac{K^{-1}-1+N_{l_{n}}}{N}\right),
\end{align*}
and $\mathfrak{Z}=1$ by Proposition \ref{prop:normConst_GF}.

An appealing aspect of $\varrho_{n}$ is that
it can be seen as the linear combination of the estimators
\begin{align*}
\varrho^{\left(0\right)} &= K^{-1},\qquad  \varrho_{n}^{\left(1\right)}= \frac{1}{N-1}\sum_{m=1,m\neq n}^{N}\mathds{1}\left(l_{m}=l_{n}\right).
\end{align*}
Whilst $\varrho^{\left(0\right)}$ is just the uniform distribution
over all labels, $\varrho_{n}^{\left(1\right)}$ represents the proportion
of labels (excluding $l_{n}$) that are equal to $l_{n}$. 
\item Consider $\psi_{n}\left(x\right)\equiv\bar{q}\left(x,l_{n}\right)$
and $\varrho_{n}=\left(K^{-1}+\sum_{m=1,m\neq n}^{N}\mathds{1}\left(l_{m}=l_{n}\right)\right)/N$
for all $n\in\left\llbracket 1,N\right\rrbracket $. For this case
we use the same $\varrho_{n}$ as in the previous example, but we
consider a non-constant set $\Psi_{N}$. Recall that in balance heuristic $\psi_{n}\left(x\right)=q_{l_{n}}\left(x\right)$,
but since we are unable to evaluate the conditional distribution using
the joint $\bar{q}$ may not seem a terrible choice. We then obtain
\begin{align*}
\widehat{Z}_{GF_{2}} & =\sum_{n=1}^{N}\frac{\tilde{\pi}\left(x_{n}\right)}{\sum_{m=1}^{N}\bar{q}\left(x_{n},l_{m}\right)}\left(\frac{K^{-1}-1+N_{l_{n}}}{N}\right),
\end{align*}
which may result in $\mathfrak{Z}\neq1$ for non-trivial choices. Nevertheless, the resulting bias vanishes as $N\rightarrow\infty$ due to the consistency of $\varrho_{n}$.
\end{itemize}
\begin{example}[Running example continued]
Figure \ref{fig:Fig6} shows results for the estimators $\widehat{Z}_{GF_{1}}$
and $\widehat{Z}_{GF_{2}}$, along with those from $\widehat{Z}_{BH}$
and $\widehat{Z}_{comb}$ for comparison purposes. Observe that overall, when the proposal is concentrated (middle
boxplots), $\widehat{Z}_{GF_{1}}$ and $\widehat{Z}_{GF_{2}}$ seem
to perform better than $\widehat{Z}_{comb}$. This behaviour seems
evident even for large $K$, as shown in Subfigures (b) and (c), where
$\widehat{Z}_{comb}$ is less variable but much more biased. When
the proposal is more spread out (right boxplots), $\widehat{Z}_{GF_{2}}$
appears to have better performance than $\widehat{Z}_{GF_{1}}$ despite
the fact that in the former $\mathfrak{Z}\neq1$, i.e. it is biased
by construction. When $K$ is very large estimating accurately $Z$
remains a challenge, as shown e.g. in Subfigure (c).
\begin{figure}[!ht]
\centering
\begin{subfigure}{\linewidth}
\centering
\includegraphics[scale=0.7]{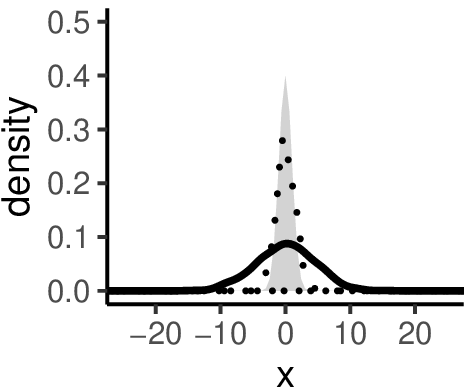}
\includegraphics[scale=0.7]{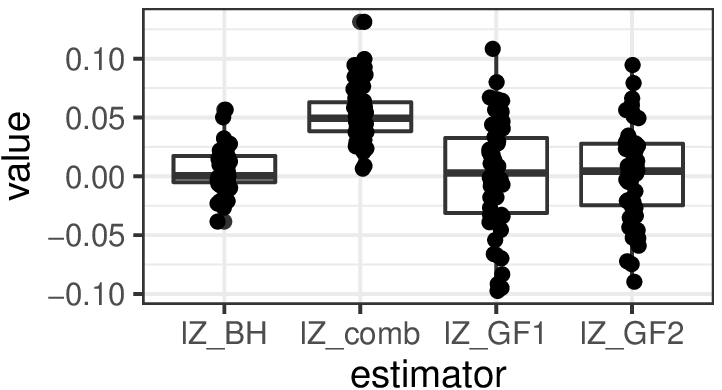}
\includegraphics[scale=0.7]{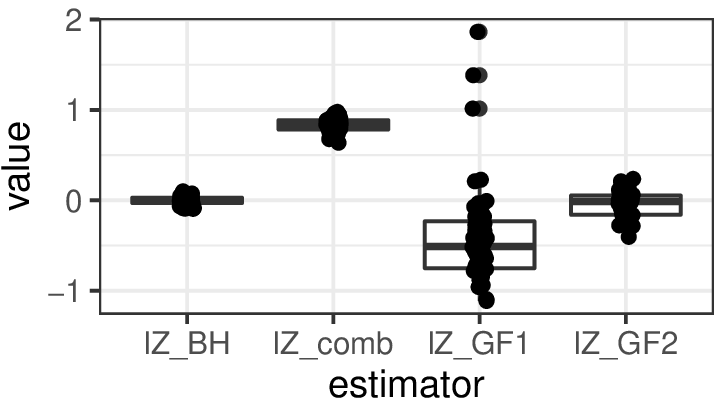}
\caption{ $K=3,000$.}
\end{subfigure}

\begin{subfigure}{\linewidth}
\centering
\includegraphics[scale=0.7]{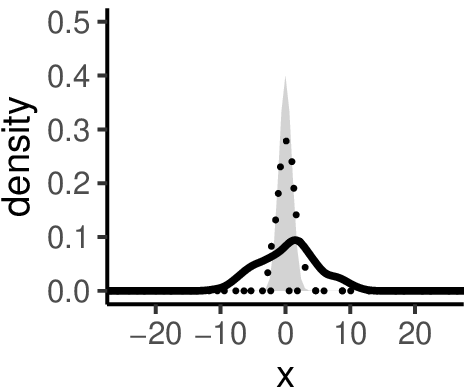}
\includegraphics[scale=0.7]{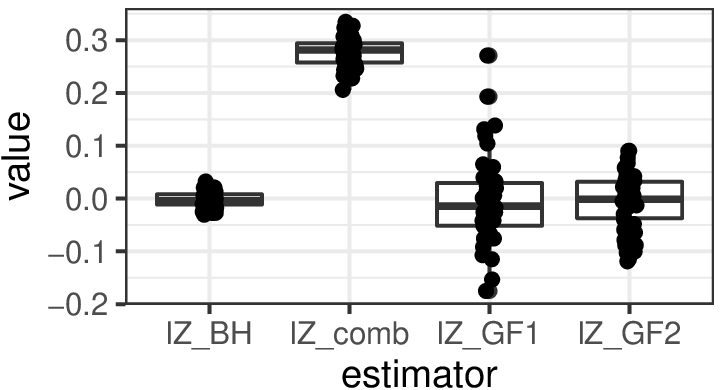}
\includegraphics[scale=0.7]{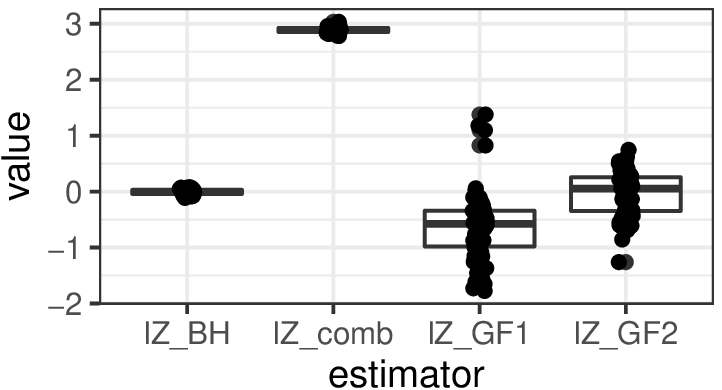}
\caption{ $K=30,000$.}
\end{subfigure}

\begin{subfigure}{\linewidth}
\centering
\includegraphics[scale=0.7]{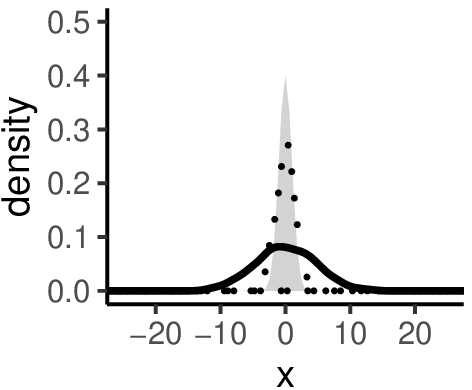}
\includegraphics[scale=0.7]{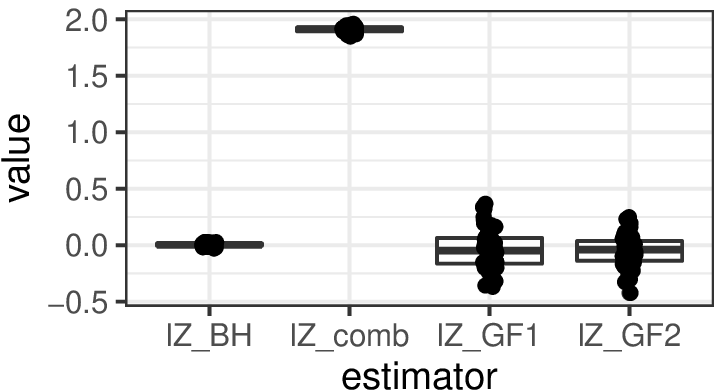}
\includegraphics[scale=0.7]{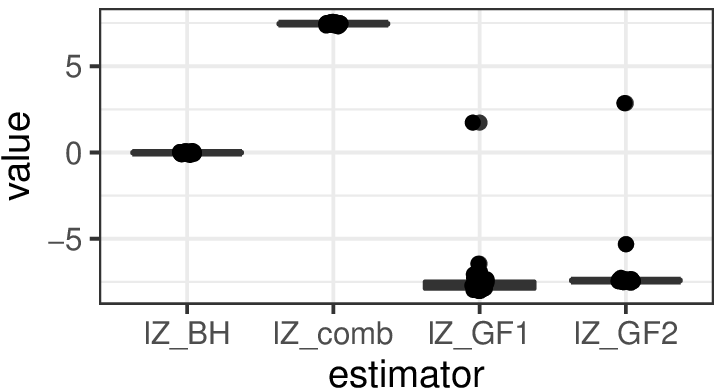}
\caption{ $K=3\times 10^6$.}
\end{subfigure}

\caption{\label{fig:Fig6}Left: true target density (shaded area) and proposal density from a sample $N=500$ for $s=\infty$ (dotted line) and $s=20$ (solid line). Middle and right: boxplots comparing normalising constant estimates in log-scale for different values of $K$ and $s=\infty$ (middle) $s=20$ (right). All cases consider $m=0.5$}
\end{figure}
\end{example}

We conclude noting that an implementation of annealed importance sampling for the general framework is also possible. For the interested reader, the details and an example can be found in Appendix 3.

\section{Discussion\label{sec:Conclusion}}

The extended representation of the balance heuristic estimator was the basis
for introducing the modified annealed importance sampling process which, in contrast to a standard
annealed importance sampling procedure, has some appealing properties that can lead to simpler
and parallel implementations. The novel representation for balance heuristic also
made clear the connection with the schemes in Section \ref{sec:intractable_proposals},
where the pool of proposals is no longer directly available.

From the various examples presented, we observe that balance heuristic (which can
be improved with the modified annealed importance sampling procedure) has consistently the
best performance for equivalent computational cost. However, when
dealing with the intractable setting from Section \ref{sec:intractable_proposals},
all the proposed estimators seem to work best whenever the marginal
proposal for the variable of interest is concentrated, meaning that
$K_{eff}\ll K$. In particular, the biased estimator $\widehat{Z}_{comb}$
that combines unbiased estimators, appears to work best if $N\approx K$
limiting its usefulness. However, the estimators $\widehat{Z}_{GF_{1}}$
and $\widehat{Z}_{GF_{2}}$ that try to mimic balance heuristic within the intractable
restrictions, can deal with more extreme cases, provided again that
$K_{eff}\ll K$. Interestingly $\widehat{Z}_{GF_{2}}$, which is biased
by construction, seems to outperform the other estimators in this
scenario. Nevertheless, the method still
struggles when the marginal distribution for $X$ is more spread out
and for very large values of $K$, even after annealed importance sampling steps.

Finally, it is worth recalling that the extended target associated
to balance heuristic does not admit $\pi$ as its marginal, however their normalising
constants are the same. Moreover and despite the previous note, the
balance heuristic (and related approximations) could be used for estimating also
expectations of functionals $f:\mathcal{X}\rightarrow\mathbb{R}$
under the target $\pi$. This can be done, when working on the extended-space
representation, by restricting to specific functions $F:\mathcal{X}^{N}\rightarrow\mathbb{R}$
of the from $F\left(x_{1:N}\right)=\sum_{n=1}^{N}f\left(x_{n}\right)$.
In terms of further improvement, it seems that the requirement $K_{eff}\ll K$
is vital for obtaining good estimates; since the annealed importance sampling procedures can
be thought of a single importance point consisting of a particle system,
it would interesting to explore how to discard bad particles without
breaking the unbiasedness. Furthermore, it seems plausible to introduce
several particle systems for implementing a sequential Monte Carlo approach, i.e. there
will be a system of particles on a higher level made of particle systems,
with the caveat of distributing the computational resources
between both levels.

\section*{Acknowledgements}
Both authors were supported by the UK Biotechnology and Biological Sciences Research Council grant BB/N00874X/1 and are grateful to Axel Finke and Chris Nemeth for useful discussions.

\bibliographystyle{plainnat}
\bibliography{MIS_V1}

\begin{thebibliography}{37}
\expandafter\ifx\csname natexlab\endcsname\relax\def\natexlab#1{#1}\fi

\bibitem[{Andrieu \& Roberts(2009)}]{AndrieuNRoberts_2009}
\textsc{Andrieu, C.} \& \textsc{Roberts, G.~O.} (2009).
\newblock {T}he pseudo-marginal approach for efficient {M}onte {C}arlo
  computations.
\newblock \textit{Ann. Statist.} \textbf{37}, 697--725.

\bibitem[{Beaumont(2003)}]{Beaumont_2003}
\textsc{Beaumont, M.~A.} (2003).
\newblock {E}stimation of {P}opulation {G}rowth or {D}ecline in {G}enetically
  {M}onitored {P}opulations.
\newblock \textit{Genetics} \textbf{164}, 1139--1160.

\bibitem[{Borsos et~al.(2019)Borsos, Curi, Levy \& Krause}]{BorsosETal_2019}
\textsc{Borsos, Z.}, \textsc{Curi, S.}, \textsc{Levy, K.~Y.} \& \textsc{Krause,
  A.} (2019).
\newblock Online variance reduction with mixtures.
\newblock \textit{arXiv preprint arXiv:1903.12416} .

\bibitem[{Capp{\'e} et~al.(2004)Capp{\'e}, Guillin, Marin \&
  Robert}]{CappeETal_2004}
\textsc{Capp{\'e}, O.}, \textsc{Guillin, A.}, \textsc{Marin, J.-M.} \&
  \textsc{Robert, C.~P.} (2004).
\newblock Population monte carlo.
\newblock \textit{Journal of Computational and Graphical Statistics}
  \textbf{13}, 907--929.

\bibitem[{Cornuet et~al.(2012)Cornuet, Marin, Mira \&
  Robert}]{CornuetETal_2012}
\textsc{Cornuet, J.-M.}, \textsc{Marin, J.-M.}, \textsc{Mira, A.} \&
  \textsc{Robert, C.~P.} (2012).
\newblock Adaptive multiple importance sampling.
\newblock \textit{Scandinavian Journal of Statistics} \textbf{39}, 798--812.

\bibitem[{Del~Moral et~al.(2006)Del~Moral, Doucet \& Jasra}]{MoralETal_2006}
\textsc{Del~Moral, P.}, \textsc{Doucet, A.} \& \textsc{Jasra, A.} (2006).
\newblock {S}equential {M}onte {C}arlo samplers.
\newblock \textit{Journal of the Royal Statistical Society: Series B
  (Statistical Methodology)} \textbf{68}, 411--436.

\bibitem[{Delyon \& Portier(2018)}]{Delyon_2018}
\textsc{Delyon, B.} \& \textsc{Portier, F.} (2018).
\newblock Efficiency of adaptive importance sampling.
\newblock \textit{arXiv preprint arXiv:1806.00989} .

\bibitem[{Douc et~al.(2007)Douc, Guillin, Marin, Robert et~al.}]{DoucETal_2007}
\textsc{Douc, R.}, \textsc{Guillin, A.}, \textsc{Marin, J.-M.}, \textsc{Robert,
  C.~P.} et~al. (2007).
\newblock Convergence of adaptive mixtures of importance sampling schemes.
\newblock \textit{The Annals of Statistics} \textbf{35}, 420--448.

\bibitem[{Doucet et~al.(2001)Doucet, de~Freitas \& Gordon}]{DoucetETal_2001}
\textsc{Doucet, A.}, \textsc{de~Freitas, N.} \& \textsc{Gordon, N.} (2001).
\newblock \textit{{S}equential {M}onte {C}arlo {M}ethods in {P}ractice}.
\newblock Information Science and Statistics. Springer New York.

\bibitem[{Elvira et~al.(2015)Elvira, Martino, Luengo \&
  Bugallo}]{ElviraETal_2015}
\textsc{Elvira, V.}, \textsc{Martino, L.}, \textsc{Luengo, D.} \&
  \textsc{Bugallo, M.~F.} (2015).
\newblock Efficient multiple importance sampling estimators.
\newblock \textit{IEEE Signal Processing Letters} \textbf{22}, 1757--1761.

\bibitem[{Elvira et~al.(2017)Elvira, Martino, Luengo \&
  Bugallo}]{ElviraETal_2017}
\textsc{Elvira, V.}, \textsc{Martino, L.}, \textsc{Luengo, D.} \&
  \textsc{Bugallo, M.~F.} (2017).
\newblock Improving population monte carlo: Alternative weighting and
  resampling schemes.
\newblock \textit{Signal Processing} \textbf{131}, 77--91.

\bibitem[{Elvira et~al.(2019)Elvira, Martino, Luengo, Bugallo
  et~al.}]{ElviraETal_2019}
\textsc{Elvira, V.}, \textsc{Martino, L.}, \textsc{Luengo, D.},
  \textsc{Bugallo, M.~F.} et~al. (2019).
\newblock Generalized multiple importance sampling.
\newblock \textit{Statistical Science} \textbf{34}, 129--155.

\bibitem[{{Everitt} et~al.(2016){Everitt}, {Culliford}, {Medina-Aguayo} \&
  {Wilson}}]{EverittETal_2016}
\textsc{{Everitt}, R.~G.}, \textsc{{Culliford}, R.}, \textsc{{Medina-Aguayo},
  F.} \& \textsc{{Wilson}, D.~J.} (2016).
\newblock {Sequential Monte Carlo with transformations}.
\newblock \textit{ArXiv e-prints} .

\bibitem[{Gramacy et~al.(2010)Gramacy, Samworth \& King}]{GramacyETal_2010}
\textsc{Gramacy, R.}, \textsc{Samworth, R.} \& \textsc{King, R.} (2010).
\newblock Importance tempering.
\newblock \textit{Statistics and Computing} \textbf{20}, 1--7.

\bibitem[{He \& Owen(2014)}]{HeNOwen_2014}
\textsc{He, H.~Y.} \& \textsc{Owen, A.~B.} (2014).
\newblock Optimal mixture weights in multiple importance sampling.
\newblock \textit{arXiv preprint arXiv:1411.3954} .

\bibitem[{Hesterberg(1995)}]{Hesterberg_1995}
\textsc{Hesterberg, T.} (1995).
\newblock Weighted average importance sampling and defensive mixture
  distributions.
\newblock \textit{Technometrics} \textbf{37}, 185--194.

\bibitem[{Jarzynski(1997)}]{Jarzynski_1997}
\textsc{Jarzynski, C.} (1997).
\newblock {Nonequilibrium Equality for Free Energy Differences}.
\newblock \textit{Physical Review Letters} \textbf{78}, 2690.

\bibitem[{Jiang \& Singh(2018)}]{JiangETSingh_2018}
\textsc{Jiang, L.} \& \textsc{Singh, S.~S.} (2018).
\newblock Tracking multiple moving objects in images using markov chain monte
  carlo.
\newblock \textit{Statistics and Computing} \textbf{28}, 495--510.

\bibitem[{Jiang et~al.(2015)Jiang, Singh \& Y{\i}ld{\i}r{\i}m}]{JiangETal_2015}
\textsc{Jiang, L.}, \textsc{Singh, S.~S.} \& \textsc{Y{\i}ld{\i}r{\i}m, S.}
  (2015).
\newblock Bayesian tracking and parameter learning for non-linear multiple
  target tracking models.
\newblock \textit{IEEE Transactions on Signal Processing} \textbf{63},
  5733--5745.

\bibitem[{Karagiannis \& Andrieu(2013)}]{KaragNAndrieu_2013}
\textsc{Karagiannis, G.} \& \textsc{Andrieu, C.} (2013).
\newblock {Annealed Importance Sampling Reversible Jump Mcmc Algorithms}.
\newblock \textit{Journal of Computational and Graphical Statistics}
  \textbf{22}, 623--648.

\bibitem[{Li et~al.(2017)Li, Yi, Hoseinnezhad, Wang \& Kong}]{LiETal_2017}
\textsc{Li, S.}, \textsc{Yi, W.}, \textsc{Hoseinnezhad, R.}, \textsc{Wang, B.}
  \& \textsc{Kong, L.} (2017).
\newblock Multiobject tracking for generic observation model using labeled
  random finite sets.
\newblock \textit{IEEE Transactions on Signal Processing} \textbf{66},
  368--383.

\bibitem[{Li et~al.(2016)Li, Chen \& Tan}]{LiETal_2016}
\textsc{Li, W.}, \textsc{Chen, R.} \& \textsc{Tan, Z.} (2016).
\newblock Efficient sequential monte carlo with multiple proposals and control
  variates.
\newblock \textit{Journal of the American Statistical Association}
  \textbf{111}, 298--313.

\bibitem[{Martino et~al.(2015)Martino, Elvira, Luengo \&
  Corander}]{MartinoETal_2015}
\textsc{Martino, L.}, \textsc{Elvira, V.}, \textsc{Luengo, D.} \&
  \textsc{Corander, J.} (2015).
\newblock An adaptive population importance sampler: Learning from uncertainty.
\newblock \textit{IEEE Transactions on Signal Processing} \textbf{63},
  4422--4437.

\bibitem[{Martino et~al.(2017)Martino, Elvira, Luengo \&
  Corander}]{MartinoETal_2017}
\textsc{Martino, L.}, \textsc{Elvira, V.}, \textsc{Luengo, D.} \&
  \textsc{Corander, J.} (2017).
\newblock Layered adaptive importance sampling.
\newblock \textit{Statistics and Computing} \textbf{27}, 599--623.

\bibitem[{Neal(2001)}]{Neal_2001}
\textsc{Neal, R.~M.} (2001).
\newblock {Annealed Importance Sampling}.
\newblock \textit{Statistics and computing} \textbf{11}, 125--139.

\bibitem[{Nemeth et~al.(2017)Nemeth, Lindsten, Filippone \&
  Hensman}]{NemethETal_2017}
\textsc{Nemeth, C.}, \textsc{Lindsten, F.}, \textsc{Filippone, M.} \&
  \textsc{Hensman, J.} (2017).
\newblock Pseudo-extended markov chain monte carlo.
\newblock \textit{arXiv preprint arXiv:1708.05239} .

\bibitem[{Nguyen et~al.(2015)Nguyen, Septier, Peters \&
  Delignon}]{NguyenETal_2015}
\textsc{Nguyen, T. L.~T.}, \textsc{Septier, F.}, \textsc{Peters, G.~W.} \&
  \textsc{Delignon, Y.} (2015).
\newblock Efficient sequential monte-carlo samplers for bayesian inference.
\newblock \textit{IEEE Transactions on Signal Processing} \textbf{64},
  1305--1319.

\bibitem[{Owen \& Zhou(2000)}]{OwenNZhou_2000}
\textsc{Owen, A.} \& \textsc{Zhou, Y.} (2000).
\newblock {Safe and Effective Importance Sampling}.
\newblock \textit{Journal of the American Statistical Association} \textbf{95},
  135--143.

\bibitem[{Owen \& Zhou(2019)}]{OwenNZhou_2019}
\textsc{Owen, A.~B.} \& \textsc{Zhou, Y.} (2019).
\newblock The square root rule for adaptive importance sampling.
\newblock \textit{arXiv preprint arXiv:1901.02976} .

\bibitem[{Robert \& Casella(2013)}]{RobertNCasella_2013}
\textsc{Robert, C.} \& \textsc{Casella, G.} (2013).
\newblock \textit{{M}onte {C}arlo {S}tatistical {M}ethods}.
\newblock Springer Texts in Statistics. Springer New York.

\bibitem[{Roy \& Evangelou(2018)}]{RoyETEvangelou_2018}
\textsc{Roy, V.} \& \textsc{Evangelou, E.} (2018).
\newblock Selection of proposal distributions for generalized importance
  sampling estimators.
\newblock \textit{arXiv preprint arXiv:1805.00829} .

\bibitem[{Sbert \& Elvira(2019)}]{SbertNElvira_2019}
\textsc{Sbert, M.} \& \textsc{Elvira, V.} (2019).
\newblock Generalizing the balance heuristic estimator in multiple importance
  sampling.
\newblock \textit{arXiv preprint arXiv:1903.11908} .

\bibitem[{Sbert \& Havran(2017)}]{Sbert_2017}
\textsc{Sbert, M.} \& \textsc{Havran, V.} (2017).
\newblock Adaptive multiple importance sampling for general functions.
\newblock \textit{The Visual Computer} \textbf{33}, 845--855.

\bibitem[{Sbert et~al.(2016)Sbert, Havran \& Szirmay-Kalos}]{SbertETal_2016}
\textsc{Sbert, M.}, \textsc{Havran, V.} \& \textsc{Szirmay-Kalos, L.} (2016).
\newblock Variance analysis of multi-sample and one-sample multiple importance
  sampling.
\newblock In \textit{Computer Graphics Forum}, vol.~35. Wiley Online Library.

\bibitem[{Sbert et~al.(2018)Sbert, Havran \& Szirmay-Kalos}]{SbertETal_2018}
\textsc{Sbert, M.}, \textsc{Havran, V.} \& \textsc{Szirmay-Kalos, L.} (2018).
\newblock Multiple importance sampling revisited: breaking the bounds.
\newblock \textit{EURASIP Journal on Advances in Signal Processing}
  \textbf{2018}, 15.

\bibitem[{Veach \& Guibas(1995)}]{VeachNGuibas_1995}
\textsc{Veach, E.} \& \textsc{Guibas, L.~J.} (1995).
\newblock {Optimally Combining Sampling Techniques for Monte Carlo Rendering}.
\newblock In \textit{Proceedings of the 22Nd Annual Conference on Computer
  Graphics and Interactive Techniques}, SIGGRAPH '95. New York, NY, USA: ACM.

\bibitem[{Zanella \& Roberts(2018)}]{ZanellaNRoberts_2018}
\textsc{Zanella, G.} \& \textsc{Roberts, G.} (2018).
\newblock Scalable importance tempering and bayesian variable selection.
\newblock \textit{Journal of the Royal Statistical Society: Series B
  (Statistical Methodology)} .

\end{thebibliography}

\appendix

\section*{Appendix 1: Proofs} \label{A1:proofs}

\subsection*{\label{A1:proof_PropZ_BH}Proof of Proposition \ref{prop:Z_BH}}
\begin{proof}
By standard manipulations of sums and integrals we have

\begin{align*}
E\left(\widehat{Z}_{MIS}\mid N_{1:K}\right) & =\int_{\mathcal{X}}\sum_{i=1}^{K}\frac{1}{N_{i}}\sum_{j=1}^{N_{i}}w_{i}\left(x\right)\frac{\tilde{\pi}\left(x\right)}{q_{i}\left(x\right)}q_{i}\left(x\right)\mu\left(dx\right)\\
 & =\int_{\mathcal{X}}\sum_{i=1}^{K}w_{i}\left(x\right)\tilde{\pi}\left(x\right)\mu\left(dx\right) =\int_{\mathcal{X}}\tilde{\pi}\left(x\right)\mu\left(dx\right)\\
 & =Z,
\end{align*}
as claimed.
\end{proof}

\subsection*{\label{A2:proof_thmVars}Proof of Theorem \ref{thm:vars}}
\begin{proof}
Let $\sigma_{BH}^{2}$ and $\sigma_{RH}^{2}$ denote the variances
of $\widehat{Z}_{BH}/Z$ and $\widehat{Z}_{RB}/Z$, respectively.
Define also 
\begin{align*}
R_{L_{1:N}}\left(x\right) & =\frac{\pi\left(x\right)}{\sum_{m=1}^{N}q_{L_{m}}\left(x\right)},\quad S_{n,L_{1:N}}\left(x\right)=\frac{q_{L_{n}}\left(X\right)}{\sum_{m=1}^{N}q_{L_{m}}\left(X\right)}.
\end{align*}
We will first obtain a simple expression for $\sigma_{BH}^{2}$, using
the equivalent representation for $\widehat{Z}_{BH}$ in \eqref{eq:Z_BH_alt}
\begin{align*}
 & E\left(\left(\widehat{Z}_{BH}/Z\right)^{2}\mid L_{1:N}\right) =\sum_{n=1}^{N}E_{X_{n}\sim q_{Ln}}\left(R_{L_{1:N}}\left(X_{n}\right)\mid L_{1:K}\right)^{2} \\
 &\qquad \quad +2\sum_{k<n}E_{X_{n}\sim q_{L_{n}}}\left(R_{L_{1:N}}\left(X_{n}\right)\mid L_{1:K}\right)E_{X_{n}\sim q_{L_{n}}}\left(R_{L_{1:N}}\left(X_{n}\right)\mid L_{1:K}\right)\\
 & \quad =\sum_{n=1}^{N}E_{X\sim\pi}\left(R_{L_{1:N}}\left(X\right)S_{n,L_{1:N}}\left(X\right)\mid L_{1:K}\right) \\
 & \qquad \quad+2\sum_{k<n}\underbrace{E_{X\sim\pi}\left(S_{n,L_{1:N}}\left(X\right)\mid L_{1:K}\right)}_{\xi_{n,L_{1:K}}}E_{X\sim\pi}\left(S_{k,L_{1:N}}\left(X\right)\mid L_{1:K}\right)\\
 & \quad =E_{X\sim\pi}\left(R_{L_{1:N}}\left(X_{n}\right)\mid L_{1:K}\right)+2\sum_{k<n}\xi_{n}\xi_{k}.
\end{align*}
The variables $\left\{ \xi_{i}\left(L_{1:N}\right)\right\} _{i}$
are identically distributed and sum up to 1, hence
\begin{align*}
1 & =E\left(\sum_{j=1}^{N}\xi_{j}\right)^{2}=\sum_{j}E\left(\xi_{j}^{2}\right)+2\sum_{j<k}E\left(\xi_{j}\xi_{k}\right)\\
 & =NE\left(\xi_{1}^{2}\right)+N\left(N-1\right)E\left(\xi_{1}\xi_{2}\right),
\end{align*}
which leads to 
\begin{align*}
\sigma_{BH}^{2} & =E\left(\widehat{Z}_{BH}/Z\right)^{2}-1=E_{L}E_{X\sim\pi}\left(R_{L_{1:N}}\left(X\right)\mid L_{1:K}\right)+N\left(N-1\right)E\left(\xi_{1}\xi_{2}\right)-1\\
 & =E_{L}E_{X\sim\pi}\left(R_{L_{1:N}}\left(X\right)\mid L_{1:K}\right)-NE\left(\xi_{1}^{2}\right).
\end{align*}

Let $\psi\left(x\right)=\sum_{l}^{K}\alpha\left(l\right)q_{l}\left(x\right)$
denote the marginal density for $X$, then by properties of the arithmetic
and geometric means
\begin{align*}
 & E_{L}E_{X\sim\pi}\left(R_{L_{1:N}}\left(X\right)\mid L_{1:K}\right) =\frac{1}{N}E_{X}\left(\frac{\pi}{\psi}\left(X\right)E_{L}\left(\frac{\psi}{\frac{1}{N}\sum_{j}q_{L_{j}}}\left(X\right)\mid X\right)\right)\\
 & \quad \leq\frac{1}{N}E_{X}\left(\frac{\pi}{\psi}\left(X\right)E_{L}\left(\psi\prod_{j} q_{L_{j}}^{-1/N}\left(X\right)\mid X\right)\right) =\frac{1}{N}E_{X}\left(\frac{\pi}{\psi}\left(X\right)\left(E_{L\mid X}^N\left(\frac{\psi}{q_{L_{j}}}\left(X\right)\right)^{1/N}\right)\right).
\end{align*}
Using the mean value theorem, the inner expectation above satisfies
for some $p^{*}\in\left(0,1/N\right)$
\begin{align*}
 & E^N_{L\mid X}\left(\left(\frac{\psi}{q_{L_{j}}}\right)^{1/N}\left(X\right)\right)=\exp\left\{ N\log\left(E_{L\mid X}\left(\frac{\psi}{q_{L}}\left(X\right)\right)^{1/N}\right)\right\} \\
 & \qquad\leq\exp\left\{ E_{L\mid X}\frac{\left(\frac{\psi}{q_{L}}\left(X\right)\right)^{1/N}-1}{1/N}\right\} =\exp\left\{ p^{*}E_{L\mid X}\log\left(\frac{\psi}{q_{L}}\left(X\right)\right)\right\} ,
\end{align*}
therefore, assuming the proposals are bounded above and below by $C_{+}$
and $C_{-}$ respectively, we obtain
\begin{align*}
\sigma_{BH}^{2} & \leq\frac{1}{N}E_{X}\left(\frac{\pi}{\psi}\left(X\right)\right)\exp\left\{ \frac{1}{N}\log\left(\frac{C_{+}}{C_{-}}\right)\right\} -NE\left(\xi_{1}^{2}\right).\\
 & \leq\frac{1}{N}E_{X}\left(\frac{\pi}{\psi}\left(X\right)\right)\exp\left\{ \frac{1}{N}\log\left(\frac{C_{+}}{C_{-}}\right)\right\} -\frac{1}{N},
\end{align*}
where the last line comes from using Jensen's inequality and noting
that $E\left(\xi_{1}\right)=1/N$. The variance $\sigma_{RH}^{2}=N^{-1}E_{X\sim\pi}\left(\left(\pi/\psi\right)\left(X\right)\right)-N^{-1}$,
which implies
\begin{align*}
\sigma_{BH}^{2} & \leq\left(\sigma_{RH}^{2}+\frac{1}{N}\right)\exp\left\{ \frac{1}{N}\log\left(\frac{C_{+}}{C_{-}}\right)\right\} -\frac{1}{N}\\
 & =\sigma_{RH}^{2}+\left(\sigma_{RH}^{2}+\frac{1}{N}\right)\left(\exp\left\{ \frac{1}{N}\log\left(\frac{C_{+}}{C_{-}}\right)\right\} -1\right)\\
 & \leq\sigma_{RH}^{2}+\frac{1}{N^{2}}\left(N\sigma_{RH}^{2}+1\right)\log\left(\frac{C_{+}}{C_{-}}\right)\exp\left\{ \frac{1}{N}\log\left(\frac{C_{+}}{C_{-}}\right)\right\} .
\end{align*}

For the result on the lower bound, by Jensen's inequality and the
relationship of the arithmetic and geometric means
\begin{align*}
\sigma_{BH}^{2} & \geq E_{X}\left(\frac{\pi\left(X\right)}{E_{L}\left(\sum_{m=1}^{N}q_{L_{m}}\left(X\right)\mid X\right)}\right)-NE\left(\xi_{1}^{2}\right)\\
 & \geq\frac{1}{N}E_{X}\left(\frac{\pi\left(X\right)}{\psi\left(X\right)}\right)-NE_{\pi}\left[\frac{1}{N}\left(E_{L}\left(q_{L_{1}}\left(X\right)\prod_{j}q_{L_{J}}^{-1/N}\left(X\right)\mid X\right)\right)^{2}\right]\\
 & =\sigma_{RB}^{2}+\frac{1}{N}-\frac{1}{N}E_{\pi}\left[\left(\frac{E_{L_{1}\mid X}q_{L_{1}}^{1-1/N}\left(X\right)}{E_{L_{2}\mid X}q_{L_{2}}^{-1/N}\left(X\right)}\left(E_{L_{2}\mid X}q_{L_{2}}^{-1/N}\left(X\right)\right)^{N}\right)^{2}\right]\\
 & \geq\sigma_{RB}^{2}-\frac{1}{N}E_{\pi}\left[\left(E_{L_{1}\mid X}q_{L_{1}}\left(X\right)\left(E_{L_{2}\mid X}q_{L_{2}}^{-1/N}\left(X\right)\right)^{N}\right)^{2}-1\right],
\end{align*}
where the last inequality comes from applying Jensen's and H\"{o}lder's
inequalities in that order to the ratio of expectations. Similarly
as before, under the assumptions for the proposals,
\begin{align*}
E^N_{L_{2}\mid X}q_{L_{2}}^{-1/N}\left(X\right) & \leq\exp\left\{ \frac{1}{N}E_{L}\left|\log\left(q_{L}\right)\right|\right\} \\
 & \leq\exp\left\{ \frac{1}{N}\log\left(\max\left\{ C_{-}^{-1},C_{+}\right\} \right)\right\} ,
\end{align*}
obtaining 
\begin{align*}
\sigma_{BH}^{2} & \geq\sigma_{RB}^{2}-\frac{1}{N}\left(\exp\left\{ \frac{2}{N}\log\left(\max\left\{ C_{-}^{-1},C_{+}\right\} \right)\right\} E_{\pi}\left[\left(E_{L_{1}\mid X}q_{L_{1}}\left(X\right)\right)^{2}\right]-1\right)\\
 & \geq\sigma_{RB}^{2}-\frac{1}{N}\left(C_{+}^{2}\exp\left\{ \frac{2}{N}\log\left(\max\left\{ C_{-}^{-1},C_{+}\right\} \right)\right\} -1\right),
\end{align*}
as claimed.
\end{proof}

\subsection*{\label{A2:proof_thmModAIS}Proof of Theorem \ref{thm:modAIS}}
\begin{proof}
First, notice that the variables $L_{1:N}$ remain unchanged throughout
the process since $\mathcal{K}_{t,n}$ only acts on the variable $\theta_{t}\left[X_{n}\right]$.
Hence, for either geometric scheme 
\begin{align*}
\prod_{t=1}^{T}\tilde{W}_{t}^{\left(n\right)} & =\frac{\tilde{\eta}_{1}\left(n,\theta_{0}\right)}{\bar{q}^{\otimes N}\left(\theta_{0}\right)}\prod_{t=1}^{T-1}\frac{\tilde{\eta}_{t+1}\left(n,\theta_{t}\right)}{\tilde{\eta}_{t}\left(n,\theta_{t}\right)}\\
 & =\frac{\tilde{\eta}_{1}\left(\theta_{0}\left[x_{n}\right]\mid n,l_{1:N}\right)}{q_{l_{n}}\left(\theta_{0}\left[x_{n}\right]\right)}\prod_{t=1}^{T-1}\frac{\tilde{\eta}_{t+1}\left(\theta_{t}\left[x_{n}\right]\mid n,l_{1:N}\right)}{\tilde{\eta}_{t}\left(\theta_{t}\left[x_{n}\right]\mid n,l_{1:N}\right)}.
\end{align*}
This reveals the true nature of the algorithm where, for each $n\in\left\llbracket 1,N\right\rrbracket $,
an independent annealed importance sampling process is carried out on the sequence of targets
$\left\{ \tilde{\eta}_{t}\left(dx_{n}\mid n,l_{1:N}\right)\right\} _{t=1}^{T}$.
Since $\mathcal{K}_{t,n}$ is an $\eta_{t}\left(dx_{n}\mid n,l_{1:N}\right)$-invariant
kernel we have that
\begin{align*}
E\left(\prod_{t=1}^{T}\tilde{W}_{t}^{\left(n\right)}\mid L_{1:N}=l_{1:N}\right) & =Z_{1,n}\prod_{t=1}^{T-1}\frac{Z_{t+1,n}}{Z_{t,n}},
\end{align*}
where $Z_{t,n}=\int_{\mathcal{X}}\tilde{\eta}_{t}\left(x\mid n,l_{1:N}\right)\mu\left(dx\right)$.
Finally, notice that 
\begin{align*}
Z_{T,n} & =\int_{\mathcal{X}}\frac{\tilde{\pi}\left(x_{n}\right)q_{l_{n}}\left(x_{n}\right)}{\sum_{m=1}^{N}q_{l_{m}}\left(x_{n}\right)}\mu\left(dx\right) = Z E_{X\sim\pi}\left(\frac{q_{l_{n}}\left(x_{n}\right)}{\sum_{m=1}^{N}q_{l_{m}}\left(x_{n}\right)}\right),
\end{align*}
which implies
\begin{align*}
E\left(\widehat{Z}\mid L_{1:N}\right) & =\sum_{n=1}^{N}E\left(\prod_{t=1}^{T}\tilde{W}_{t}^{\left(n\right)}\mid L_{1:N}\right)\\
 & =Z\sum_{n=1}^{N} E_{X\sim\pi}\left(\frac{q_{L_{n}}\left(x_{n}\right)}{\sum_{m=1}^{N}q_{L_{m}}\left(x_{n}\right)}\mid L_{1:N}\right)\\
 & =Z,
\end{align*}
as required.
\end{proof}

\subsection*{\label{A3:proof_lemMultEstims}Proof of Lemma \ref{lem:multEstims}}
\begin{proof}
By the Tower property
\begin{align*}
E\left(\widehat{Z}_{i}\right) & =\frac{1}{N}E\left(N_{i}\right) E_{X\sim q_{i}}\left( \frac{\tilde{\pi}\left(X\right)}{\bar{q}\left(X,i\right)}\right)\\
 & =\alpha_{i}\int\frac{\tilde{\pi}\left(x\right)}{\bar{q}\left(x,i\right)}q_{i}\left(x\right)\mu\left(dx\right)\\
 & =Z.
\end{align*}
For the second result, use the variance decomposition formula to obtain
\begin{align*}
\mbox{var}\left( \widehat{Z}_{i}\right) & =\frac{1}{N^{2}}\left(E\left(N_{i}\right) \mbox{var}_{X\sim q_{i}}\left( \frac{\tilde{\pi}\left(X\right)}{\bar{q}\left(X,i\right)}\right)+ \mbox{var}\left( N_{i}\right) E_{X\sim q_{i}}^{2}\left( \frac{\tilde{\pi}\left(X\right)}{\bar{q}\left(X,i\right)}\right)\right)\\
 & =\frac{1}{N^{2}}\left(N\alpha_{i}E\left(\frac{\tilde{\pi}\left(X_{i1}\right)}{\bar{q}\left(X_{i1},i\right)}\right)^{2}-N\alpha_{i}^{2}\left(\frac{Z}{\alpha_{i}}\right)^{2}\right)\\
 & =\frac{1}{N}\left(\int\frac{\tilde{\pi}\left(x\right)}{\bar{q}\left(x,i\right)}\tilde{\pi}\left(x\right)\mu\left(dx\right)-Z^{2}\right)\\
 & =\frac{Z^{2}}{N}\left(E_{X\sim\pi}\left(\frac{\pi\left(X\right)}{\bar{q}\left(X,i\right)}\right)-1\right).
\end{align*}
Finally, for the covariance expression, if $i\neq j$ 
\begin{align*}
\mbox{cov}\left( \widehat{Z}_{i},\widehat{Z}_{j}\right) & =\frac{1}{N^{2}}E\left(N_{i}N_{j}\right) E_{X\sim q_{i}}\left( \frac{\tilde{\pi}\left(X\right)}{\bar{q}\left(X,i\right)}\right) E_{X\sim q_{j}}\left(\frac{\tilde{\pi}\left(X\right)}{\bar{q}\left(X,j\right)}\right)-Z^{2}\\
 & =\frac{1}{N^{2}}\left(\mbox{cov}\left( N_{i},N_{j}\right) +N^{2}\alpha_{i}\alpha_{j}\right)\frac{Z^{2}}{\alpha_{i}\alpha_{j}}-Z^{2}\\
 & =\frac{N\left(N-1\right)}{N^{2}}Z^{2}-Z^{2}\\
 & =-\frac{Z^{2}}{N},
\end{align*}
as claimed.
\end{proof}

\subsection*{\label{A4: proof_thmOptWeights}Proof of Theorem \ref{thm:optimalWeights}}
\begin{proof}
The variance of $\widehat{Z}_{comb}$ is
\begin{align*}
\mbox{var} \left(\widehat{Z}_{comb}\right) & =\mbox{var}\left(\nu^T_{1:K}\widehat{Z}_{1:K}\right)\\
 & =\nu^T_{1:K}\Sigma_{K}\nu{}_{1:K}.
\end{align*}
Using Lagrange multipliers we aim at minimising the function 
\begin{align*}
F\left(\nu_{1:K},\lambda\right) & =\nu^T_{1:K}\Sigma_{K}\nu{}_{1:K}-\lambda\left(\mathbf{1}^T_{1:K}\nu{}_{1:K}-1\right).
\end{align*}
By matrix differentiation
\begin{align*}
\frac{\partial F}{\partial\nu_{1:K}} & =2\nu^T_{1:K}\Sigma_{K}-\lambda\mathbf{1}^T_{1:K}\quad\text{and}\quad\frac{\partial F}{\partial\lambda}=1-\mathbf{1}^T_{1:K}\nu{}_{1:K};
\end{align*}
hence, the only critical point is such that $\nu{}_{1:K}=0.5\lambda\Sigma_{K}^{-1}\mathbf{1}{}_{1:K}$
and $\lambda^{-1}=0.5 \mathbf{1}^T_{1:K}\Sigma_{K}^{-1}\mathbf{1}{}_{1:K}$.
Since $\Sigma_{K}$ is a positive definite matrix the critical point
must be the global minimum. Therefore
\begin{align*}
\nu^{opt}{}_{1:K} & =\frac{1}{2}\lambda^{opt}\Sigma_{K}^{-1}\mathbf{1}{}_{1:K} =\frac{\Sigma_{K}^{-1}\mathbf{1}_{1:K}}{\mathbf{1}'_{1:K}\Sigma_{K}^{-1}\mathbf{1}_{1:K}},
\end{align*}
which completes the proof.
\end{proof}

\subsection*{\label{A5:proof_propNormConst_GF}Proof of Proposition \ref{prop:normConst_GF}}
\begin{proof}
Using \eqref{eq:eta_new}, we first integrate out the vector $x_{1:N}$
leading to 
\[
E_{X\sim\pi}\left( \frac{\psi_{n}\left(X\right)}{\sum_{m=1}^{N}\psi_{m}\left(X\right)}\right) \frac{\varrho_{n}}{\alpha\left(l_{n}\right)}\alpha^{\otimes N}\left(l_{1:N}\right).
\]
Taking the sum over $n\in\left[1:N\right]$ and summing over the vector
$l_{1:N}$ we obtain the first part of the result.
For the second part, if $\psi\equiv 1$
\begin{align*}
\mathfrak{Z} & =E_{L_{1:N}\sim\alpha^{\otimes N}}\left( \frac{1}{N}\sum_{n=1}^{N}\frac{\varrho_{n}}{\alpha\left(L_{n}\right)}\right)=\frac{1}{N}E\left(\sum_{n=1}^{N}\frac{K^{-1}+N_{l_{n}}-1}{N\alpha\left(L_{n}\right)}\right)\\
 & =\frac{1}{N}E\left(\sum_{j=1}^{K}\sum_{i=1}^{N_{j}}\frac{K^{-1}+N_{j}-1}{N\alpha_{j}}\right)=\frac{1}{N}E\left(\sum_{j=1}^{K}N_{j}\frac{K^{-1}+N_{j}-1}{N\alpha_{j}}\right)\\
 & =\frac{1}{N}\sum_{j=1}^{K}\frac{N\alpha_{j}\left(K^{-1}-1\right)+E\left(N_{j}^{2}\right]}{N\alpha_{j}}=\frac{1}{N}\sum_{j=1}^{K}\left[K^{-1}-\alpha_{j}+N\alpha_{j}\right] =1,
\end{align*}
as claimed.
\end{proof}

\section*{Appendix 2: Conceptual example}

Suppose we have available an ordered sample $Y_{1:n}$ with joint
distribution $\pi_{n}$ and support 
\[
\mathcal{X}^{n}=\left\{ x_{1:n}\in\mathbb{R}^{n}\mid x_{1}<\dots<x_{n}\right\} .
\]
We want to use IS for estimating some joint distribution $\pi_{n+1}$
on $\mathcal{X}^{n+1}$ using the ordered sample $Y_{1:n}$. A simple
approach would be to sample $\tilde{Y}_{n+1}\sim\varphi\left(\cdot\right)$,
from some distribution $\varphi$ on $\left(\mathbb{R},\mathcal{B}\left(\mathbb{R}\right)\right)$,
and then transform $\left(Y_{1:n},\tilde{Y}_{n+1}\right)$ in such
way that we obtain a new ordered sample $X_{1:n+1}$. Simply letting
$X_{1:n+1}=\left(Y_{1:n},\tilde{Y}_{n+1}\right)$ could be problematic
(leading to IS weights that could be exactly zero) unless $\tilde{Y}_{n+1}>Y_{n}$
with probability one; however, notice that a simple reordering of the vector $\left(Y_{1:n},\tilde{Y}_{n+1}\right)$
will guarantee that $\pi_{n+1}\left(x_{1:n+1}\right)>0$, i.e. we
set $X_{1:n+1}=\mbox{{order}}\left(Y_{1:n},\tilde{Y}_{n+1}\right)$.
By doing this, we have implicitly introduced a discrete variable $L\in\left\llbracket 1, n+1 \right\rrbracket$
indicating the position of the last sampled variable $\tilde{Y}_{n+1}$
within $X_{1:n+1}$, meaning that $X_{L}=\tilde{Y}_{n+1}$ and $X_{-L}=Y_{1:n}$.
The introduction of this variable is essential for obtaining the density
of $X_{1:n+1}$, which is a one of the key ingredients of the IS estimate.
The joint density for $\left(X_{1:n+1},L\right)$ is easily obtained
\begin{align*}
\bar{q}\left(x_{1:n+1},l\right) & =\pi_{n}\left(x_{-l}\right)\varphi\left(x_{l}\right).
\end{align*}
Therefore, the marginal proposal density for $X_{1:n+1}$ is obtained
by summing through all possible values for $l$
\begin{align*}
q\left(x_{1:n+1}\right) & =\sum_{l=1}^{n+1}\bar{q}\left(x_{1:n+1},l\right).
\end{align*}
Notice, however, that neither the marginal $q\left(l\right)$ nor
the conditional $q\left(x_{1:n+1}\mid l\right)$ are analytically
available in general. Thus, any IS estimate can only be obtained through
the extended proposal $\bar{q}$ or through the marginal $q\left(x_{1:n+1}\right)$,
which could be expensive to compute as $n$ increases.

\section*{Appendix 3: Annealed importance sampling for $\eta_{GF}$}

Using e.g. the
semi-geometric scheme, we obtain the following sequence of targets
for $t\in\left\llbracket 1,T\right\rrbracket $
\begin{align*}
\eta_{GF,t}^{(sg)}\left(n,x_{1:N},l_{1:N}\right) & \propto\pi\left(x_{n}\right)^{\gamma_{t}}\frac{\psi_{n}\left(x_{n}\right)^{\gamma_{t}}\bar{q}\left(x_{n},l_{n}\right)^{1-\gamma_{t}}}{\sum_{m=1}^{N}\psi_{m}\left(x_{n}\right)^{\gamma_{T}}}\varrho_{n}^{\gamma_{t}}\prod_{m\neq n}\left\{ \bar{q}\left(x_{m},l_{m}\right)\right\} .
\end{align*}
Similarly as in the case for balance heuristic, the full conditional of $x_{n}$
does not depend on the remaining variables $x_{-n}$, i.e.
\begin{align*}
\eta_{GF,t}^{(sg)}\left(x_{n}\mid n,x_{-n},l_{1:N}\right) & \propto\tilde{\pi}\left(x_{n}\right)^{\gamma_{t}}\frac{\psi_{n}\left(x_{n}\right)^{\gamma_{t}}\bar{q}\left(x_{n},l_{n}\right)^{1-\gamma_{t}}}{\sum_{m=1}^{N}\psi_{m}\left(x_{n}\right)^{\gamma_{T}}} =\tilde{\eta}_{GF,t}^{(sg)}\left(x_{n}\mid n,l_{1:N}\right).
\end{align*}
Therefore, Algorithm \ref{alg:mod_AIS} can also be implemented by
choosing $\mathcal{K}_{t,n}$ as a $\eta_{GF,t}^{(sg)}\left(dx_{n}\mid n,l_{1:N}\right)$-invariant
kernel.
\begin{example}[Running example continued]
Figure \ref{fig:Fig7} contains the results for $\widehat{Z}_{GF_{2}}$
with the implementation of the modified annealed importance sampling presented in Algorithm
\ref{alg:mod_AIS}. Boxplots on the right column consider the balance heuristic estimator
(for comparison purposes) and the three implementations of modified annealed importance sampling using
different values of $T$. Observe that when the marginal proposal
for $X$ is more spread out, the variance of the resulting estimators
is still considerably high, whereas the balance heuristic performs very well in such
cases.

\begin{figure}[!ht]{
\centering
\begin{subfigure}{\linewidth}
\centering
\includegraphics[scale=0.7]{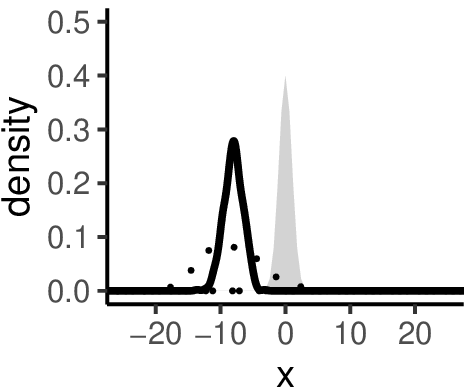}\includegraphics[scale=0.7]{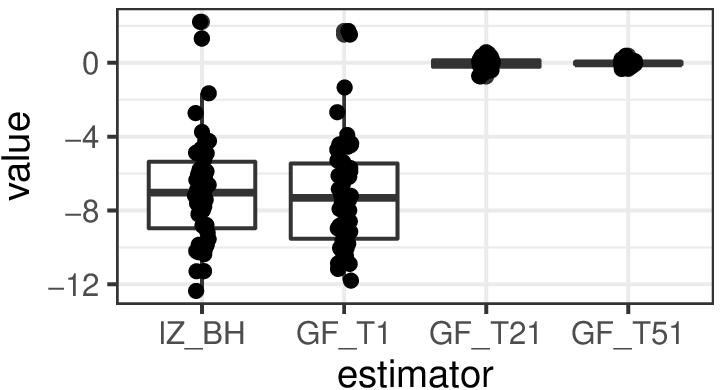}
\includegraphics[scale=0.7]{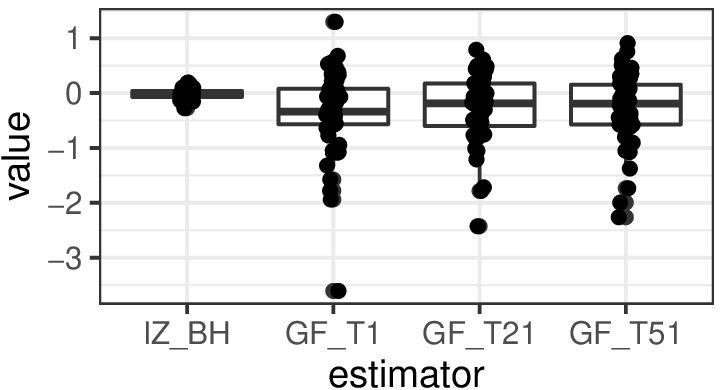}
\caption{ $K=3,000$.}
\end{subfigure}

\begin{subfigure}{\linewidth}
\centering
\includegraphics[scale=0.7]{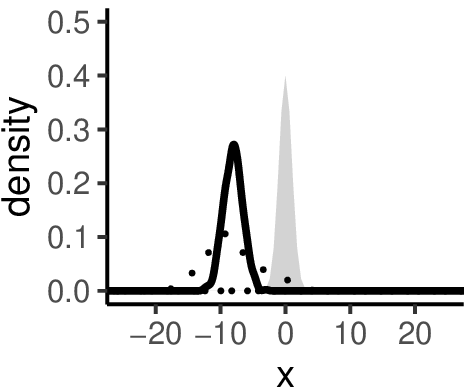}\includegraphics[scale=0.7]{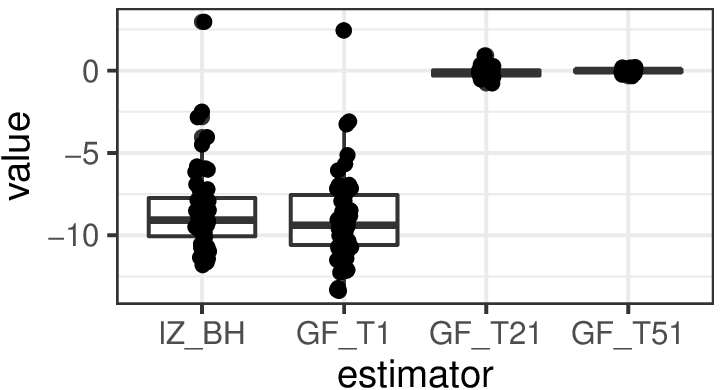}
\includegraphics[scale=0.7]{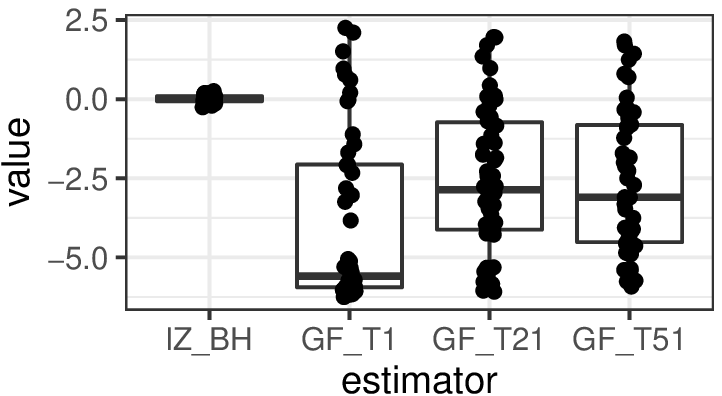}
\caption{ $K=30,000$.}
\end{subfigure}

\begin{subfigure}{\linewidth}
\centering
\includegraphics[scale=0.7]{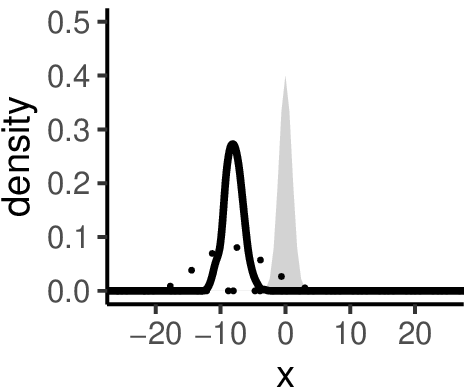}\includegraphics[scale=0.7]{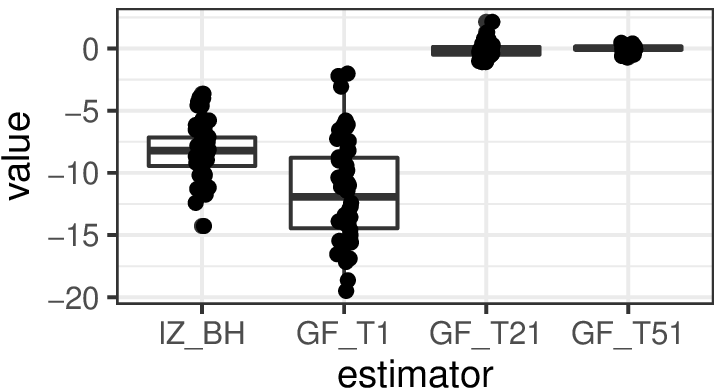}
\includegraphics[scale=0.7]{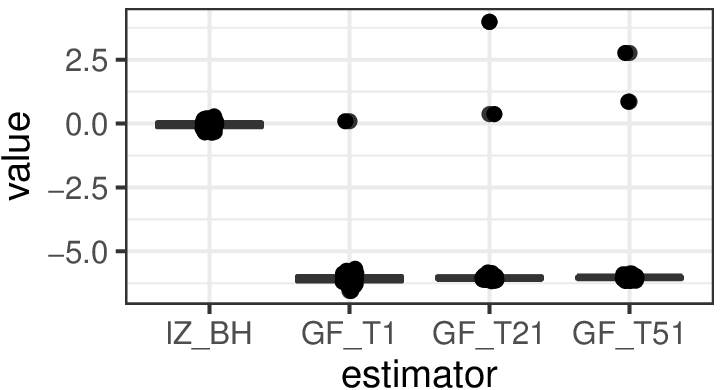}
\caption{ $K=3\times 10^6$.}
\end{subfigure}

}\caption{\label{fig:Fig7}Left: true target density (shaded area) and proposal density from a sample $N=500$ for $s=\infty$ (solid line) and $s=20$ (dotted line). Middle and right: boxplots comparing normalising constant estimates in log-scale for different values of $K$ and $s=\infty$ (middle) $s=20$ (right). All cases consider $m=0.3$}
\end{figure}
\end{example}

\section{Supplementary material}\label{D:SuppMat}

\subsection{Comparison between geometric schemes}

\begin{figure}[H]
\begin{subfigure}{\linewidth}
\includegraphics[scale=0.75]{{density_K30000_m0\lyxdot 5_s2}.eps}\includegraphics[scale=0.75]{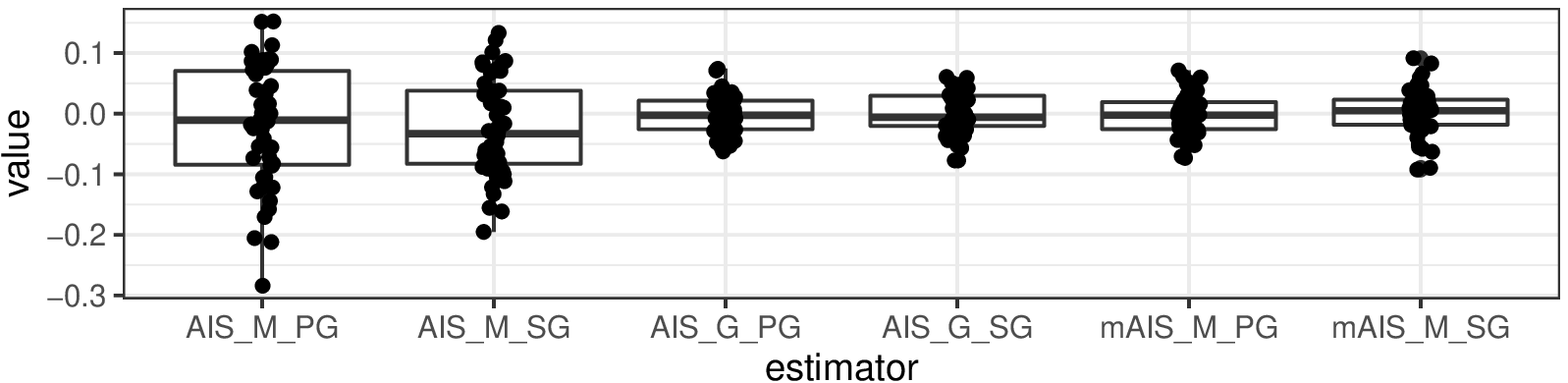}
\caption{$m=0.5, s=2$.}
\end{subfigure}

\begin{subfigure}{\linewidth}
\includegraphics[scale=0.75]{{density_K30000_m0\lyxdot 2_s20}.eps}
\includegraphics[scale=0.75]{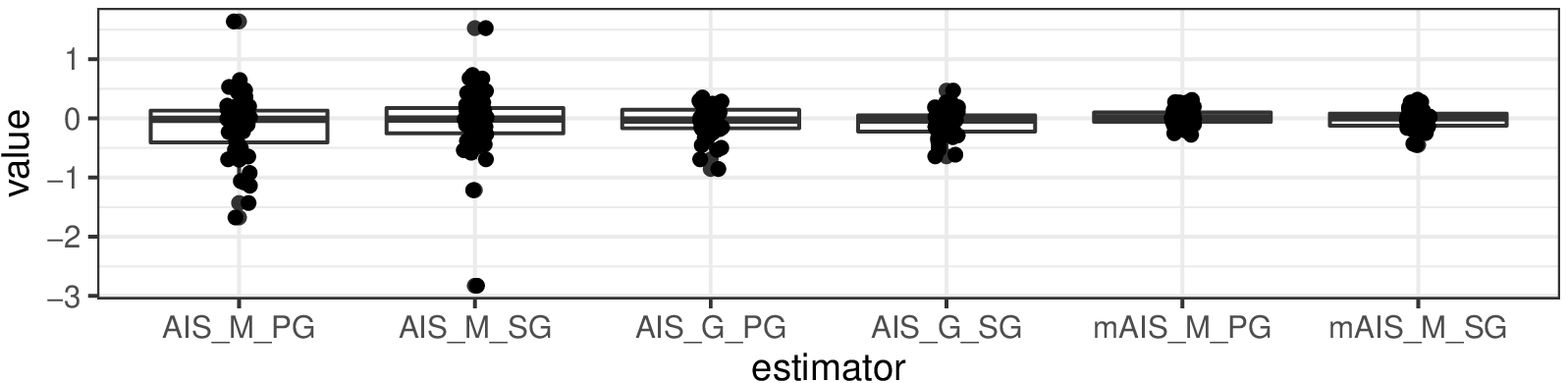}
\caption{$m=0.2, s=20$.}
\end{subfigure}

\begin{subfigure}{\linewidth}
\includegraphics[scale=0.75]{{density_K30000_m0\lyxdot 2_sInf}.eps}\includegraphics[scale=0.75]{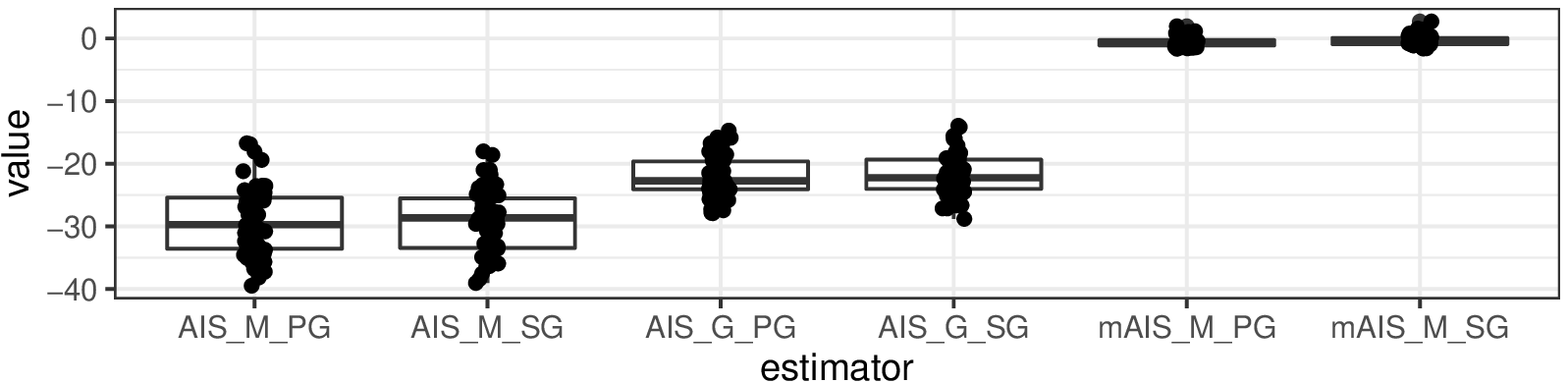}
\caption{$m=0.2, s=\infty$.}
\end{subfigure}

\caption{\label{fig:FigB1}Left: true target density (shaded area) and proposal density from
a sample $N=500$ (solid line). Right: boxplots comparing normalising constant estimates in log-scale for different values of $m$ and $s$ and between semi-geometric and purely-geometric schemes. All cases consider $K=30,000$, and $T=21$.}
\end{figure}

\subsection{Further results for $\widehat{Z}_{comb}$}

\begin{figure}[H]
\begin{subfigure}{\linewidth}
\includegraphics[scale=0.9]{{density_K500_m0\lyxdot 5_sInf}.eps}
\includegraphics[scale=0.9]{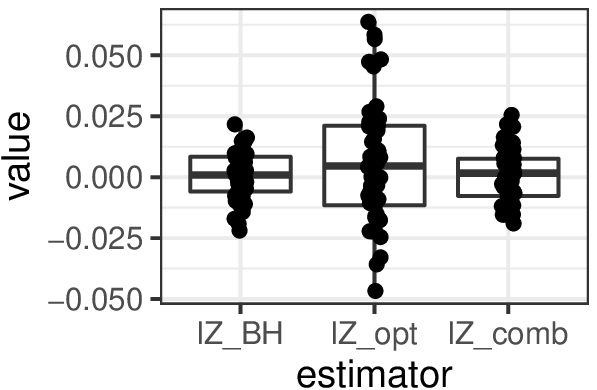}\includegraphics[scale=0.9]{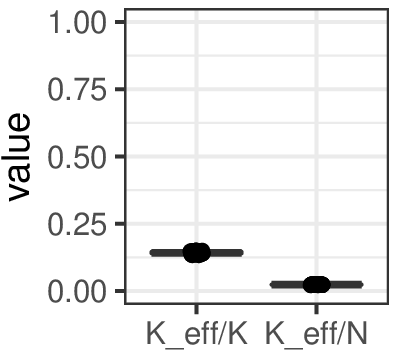}
\caption{$K=500$.}
\end{subfigure}

\begin{subfigure}{\linewidth}
\includegraphics[scale=0.9]{{density_K3000_m0\lyxdot 5_sInf}.eps}\includegraphics[scale=0.9]{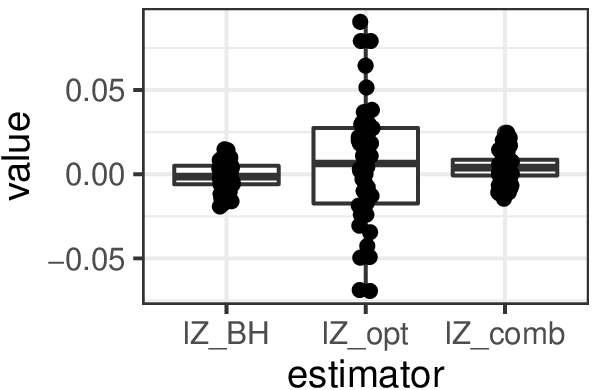}\includegraphics[scale=0.9]{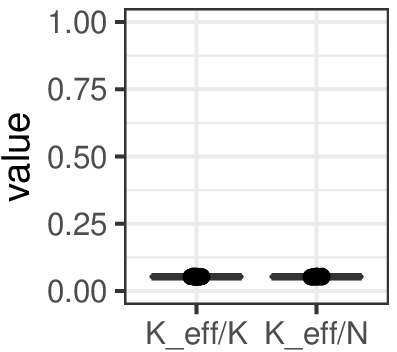}
\caption{$K=3,000$.}
\end{subfigure}

\begin{subfigure}{\linewidth}
\includegraphics[scale=0.9]{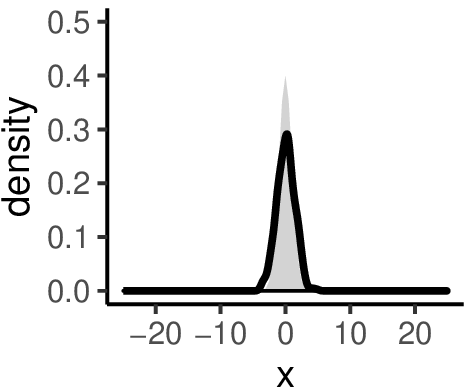}\includegraphics[scale=0.9]{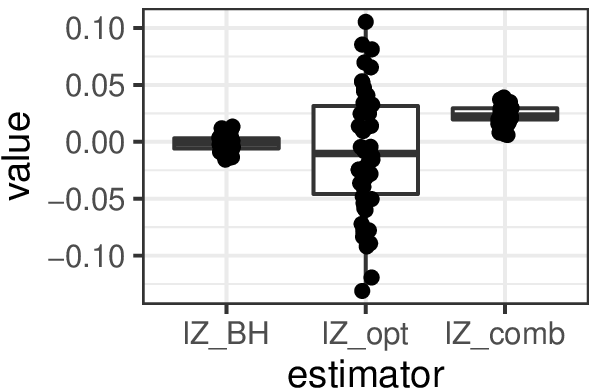}\includegraphics[scale=0.9]{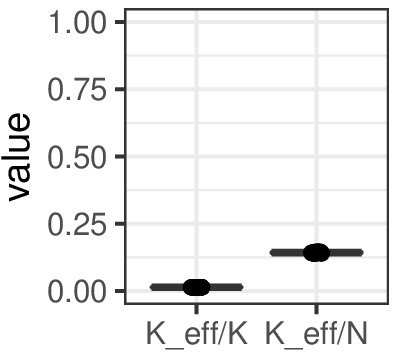}
\caption{$K=30,000$.}
\end{subfigure}

\caption{\label{fig:FigB2-1}Left: true target density (blue) and proposal density from
a sample $N=3,000$ (red). Middle: boxplots comparing normalising constant estimates in log-scale for different values of $K$. Right: boxplots for proportions $K_{eff}/K$ and $K_{eff}/N$. All cases consider $m=0.5$ and $s=\infty$.}
\end{figure}

\begin{figure}[H]
\begin{subfigure}{\linewidth}
\includegraphics[scale=0.9]{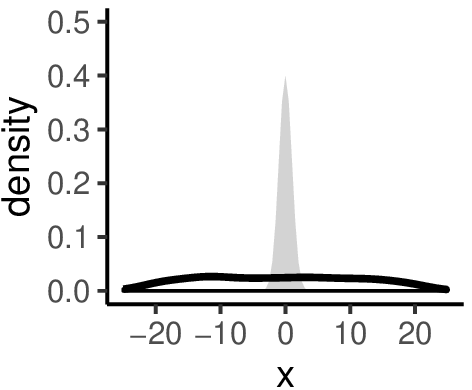}\includegraphics[scale=0.9]{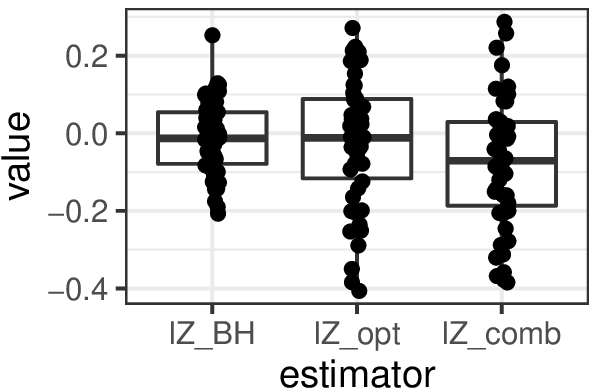}\includegraphics[scale=0.9]{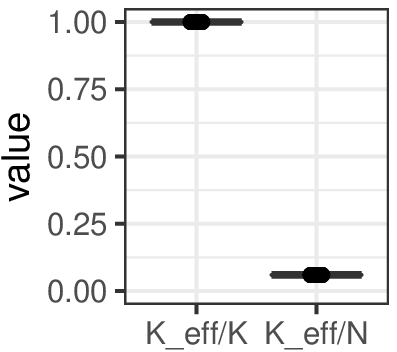}
\caption{$K=30$.}
\end{subfigure}

\begin{subfigure}{\linewidth}
\includegraphics[scale=0.9]{{density_K300_m0\lyxdot 5_s2}.eps}\includegraphics[scale=0.9]{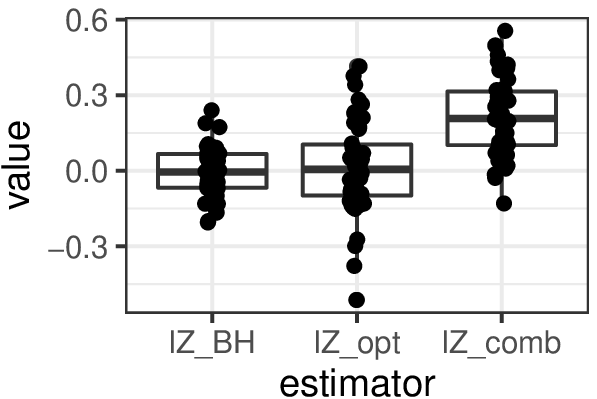}\includegraphics[scale=0.9]{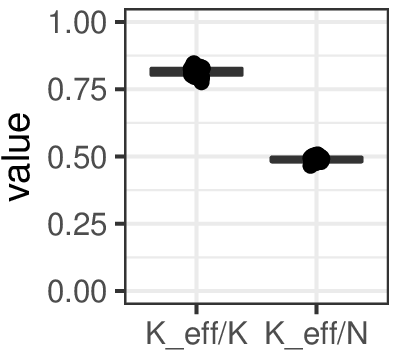}
\caption{$K=300$.}
\end{subfigure}

\begin{subfigure}{\linewidth}
\includegraphics[scale=0.9]{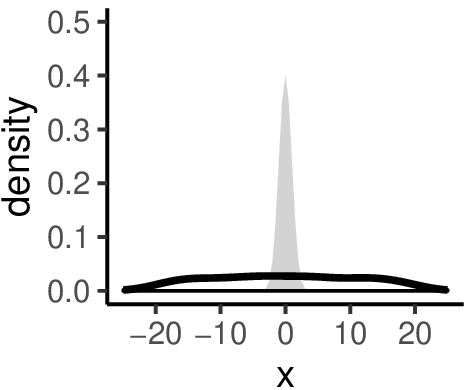}\includegraphics[scale=0.9]{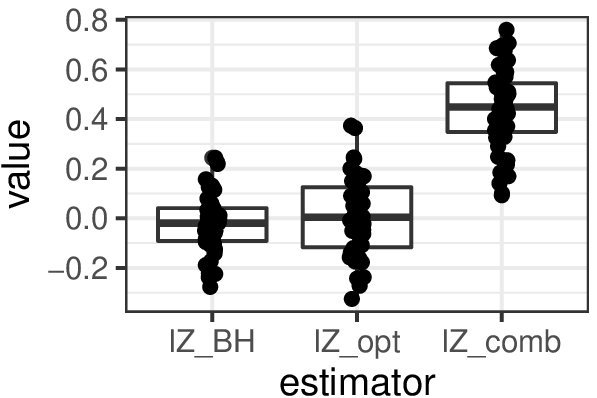}\includegraphics[scale=0.9]{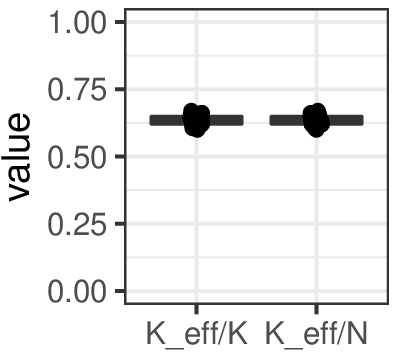}
\caption{$K=500$.}
\end{subfigure}

\caption{\label{fig:FigB2-2}Left: true target density (blue) and proposal density from
a sample $N=500$ (red). Middle: boxplots comparing normalising constant estimates in log-scale for different values of $K$. Right: boxplots for proportions $K_{eff}/K$ and $K_{eff}/N$. All cases consider $m=0.5$ and $s=2$.}
\end{figure}

\end{document}